\documentclass[10pt,a4paper,twosided,twocolumn,notitlepage]{article}

\usepackage[english]{babel}
\usepackage{graphicx}
\usepackage{dcolumn}
\usepackage{bm}

\newcommand{\IEEE}[3]      {IEEE Trans.\ Nucl.\ Sci.~{\bf NS-#1} (#2) #3}
\newcommand{\NIM}[3]       {Nucl.\ Instr.\ Methods~{\bf A#1} (#2) #3}
\newcommand{\NPA}[3]       {Nucl.\ Phys.~{\bf A#1} (#2) #3}
\newcommand{\NPB}[3]       {Nucl.\ Phys.~{\bf B#1} (#2) #3}
\newcommand{\PLB}[3]       {Phys.\ Lett.~{\bf B#1} (#2) #3}
\newcommand{\PRC}[3]       {Phys.\ Rev.~{\bf C#1} (#2) #3}
\newcommand{\PRD}[3]       {Phys.\ Rev.~{\bf D#1} (#2) #3}
\newcommand{\PRL}[3]       {Phys.\ Rev.\ Lett.~{\bf #1} (#2) #3}

\newcommand{\ZPC}[3]       {Z.~Phys.~{\bf C#1} (#2) #3}
\newcommand{\EPJ}[3]       {Eur.\ Phys.\ J.~{\bf C#1} (#2) #3}
\newcommand{\CPC}[3]       {Comp.\ Phys.\ Comm.~{\bf #1} (#2) #3}
\newcommand{\JETP}[3]       {JETP Lett.~{\bf D#1} (#2) #3}

\newcommand{\PAN}[3]       {Phys.\ Atom. Nucl.~{\bf B#1} (#2) #3}

\newdimen\Parindent\newdimen\Parskip
\Parindent=\parindent\Parskip=\parskip    
\begin{document}

\titlepage{

\flushleft{
DESY-08-180\\
\today \\
}
\vspace{3mm}
\center{\Large\bf Kinematic distributions and nuclear effects of $\boldmath{J/\psi}$ \\
production in 920 GeV fixed-target proton-nucleus collisions}
\vspace{3mm}
\center{\large\bf The HERA-B Collaboration}\\
\vspace{1mm}

I.~Abt$^{24}$,
M.~Adams$^{11}$,
M.~Agari$^{14}$,
H.~Albrecht$^{13}$,
A.~Aleksandrov$^{30}$,
V.~Amaral$^{9}$,
A.~Amorim$^{9}$,
S.~J.~Aplin$^{13}$,
V.~Aushev$^{17}$,
Y.~Bagaturia$^{13,37}$
% ...

\vspace{5mm}
\noindent
% ...
$^{9}${\it LIP Coimbra, P-3004-516 Coimbra,  Portugal~$^{f}$} \\
% ...
$^{11}${\it Institut f\"ur Physik, Universit\"at Dortmund, D-44221 Dortmund, Germany~$^{d}$} \\
% ...
$^{13}${\it DESY, D-22603 Hamburg, Germany} \\
$^{14}${\it Max-Planck-Institut f\"ur Kernphysik, D-69117 Heidelberg, Germany~$^{d}$} \\
% ...
$^{17}${\it Institute for Nuclear Research, Ukrainian Academy of Science, 03680 Kiev, Ukraine~$^{h}$} \\
% ...
$^{24}${\it Max-Planck-Institut f\"ur Physik, Werner-Heisenberg-Institut, D-80805 M\"unchen, Germany~$^{d}$} \\
% ...
$^{30}${\it DESY, D-15738 Zeuthen, Germany} \\
% ...
$^{37}${\it visitor from High Energy Physics Institute, 380086 Tbilisi, Georgia} \\
% ...
$^{d}$ supported by the Bundesministerium f\"ur Bildung und Forschung, FRG, under contract numbers 05-7BU35I, 05-7DO55P, 05-HB1HRA, 05-HB1KHA, 05-HB1PEA, 05-HB1PSA, 05-HB1VHA, 05-HB9HRA, 05-7HD15I, 05-7MP25I, 05-7SI75I \\
% ...
$^{f}$ supported by the Portuguese Funda\c c\~ao para a Ci\^encia e Tecnologia under the program POCTI \\
% ...
$^{h}$ supported by the National Academy of Science and the Ministry of Education and Science of Ukraine \\
% ...

\abstract{
Measurements of the kinematic distributions of $J/\psi$ mesons
produced in $p-$C, $p-$Ti and $p-$W collisions at
$\sqrt{s}=41.6~\mathrm{GeV}$ in the Feynman-$x$ region $-0.34 < x_{F}
< 0.14$ and for transverse momentum up to $p_T = 5.4~\mathrm{GeV}/c$
are presented.  The $x_F$ and $p_T$ dependencies of the nuclear
suppression parameter, $\alpha$, are also given. The results are based
on $2.4 \cdot 10^{5}$ $J/\psi$ mesons in both the $e^+ e^-$ and
$\mu^{+}\mu^{-}$ decay channels. The data have been collected by the
HERA-B experiment at the HERA proton ring of the DESY laboratory. The
measurement explores the negative region of $x_{F}$ for the first
time. The average value of $\alpha$ in the measured $x_{F}$ region is
$0.981 \pm 0.015$. The data suggest that the strong nuclear
suppression of $J/\psi$ production previously observed at high $x_F$
turns into an enhancement at negative $x_F$.
}\\
\vspace{3mm}

}
\twocolumn

\section{Introduction} \label{sec:introd}
The DESY experiment HERA-B has measured inclusive $J/\psi$ production
in proton-carbon, proton-titanium and proton-tungsten collisions at a
center-of-mass energy $\sqrt{s} = 41.6$\,GeV. The results are based on
a sample of about $2.4 \cdot 10^{5} \; J/\psi$ mesons reconstructed in
both dilepton decay channels. A measurement of the atomic mass number
dependence of $J/\psi$ production is derived from a comparison of the
different samples. The atomic number dependence of inclusive particle
production is often characterized by a power law: $\sigma_{pA} =
\sigma_{pN} \cdot A^\alpha$ where
$\sigma_{pN}$ is the proton-nucleon cross section and $\sigma_{pA}$ is
corresponding proton-nucleus cross section for a target of atomic mass
number $A$. Previous measurements by E866 at
Fermilab~\cite{FNAL,e866_adep} at $\sqrt{s} = 38.8$\,GeV and NA50 at
CERN~\cite{na50} at lower energy indicate that $\alpha
\sim 0.94-0.95$ at $x_{F} \sim 0$ and decreases to $\sim 0.65$ as $x_F$
approaches unity~\cite{e866_adep}. The results presented here provide
a first measurement of nuclear effects in charmonium production
extending into the negative part of the Feynman-$x$ spectrum, $-0.34 <
x_F < 0.14$
\footnote{ A slightly different range is quoted in previous HERA-B 
publications of $J/\psi$ results due to minor differences in selection
cuts.}. 

An understanding of the basic mechanisms responsible for the
suppression of charmonium production in proton-nucleus collisions
relative to proton-nucleon collisions is a prerequisite for the
identification of possible signals of new physics in high-energy
heavy-ion data. Interpretations of the existing proton-nucleus data at
positive $x_F$ rely on a delicate balance of several processes:
nuclear absorption, shadowing of parton densities, energy loss,
interactions with co-movers, hadronization of intrinsic $c\overline{c}$
components of the scattering nucleons, and so on. Ad hoc combinations
of such elementary mechanisms, considered within various theoretical
frameworks~\cite{vog_00,vog_05,borkaid} are able to qualitatively
reproduce the observed strong increase of $J/\psi$ suppression as
$x_F$ approaches unity~\cite{e866_adep}.

However the presently available data give little guidance in the
largely unexplored negative-$x_F$ region where other mechanisms such
as formation-time effects can influence the effective nuclear path
length of produced states~\cite{gavin_mi,brodsky_h,zakar,karz_satz}
and can potentially lead to a decidedly different behavior of
$\alpha$. In contrast to $J/\psi$ production in the positive $x_F$
region, at negative $x_{F}$ the produced $c\bar{c}$ pair
preferentially evolves into a charmonium state before leaving the
nucleus and nuclear effects influencing the $J/\psi$ itself become
important.  Especially in this region, different models and approaches
lead to contrasting predictions, for example arising from differing
assumptions on energy loss of beam partons or produced $c\bar{c}$
pairs.  By extending the current experimental knowledge of the nuclear
dependence of $J/\psi$ production towards negative $x_F$, the
measurement described here provides new constraints for possible
explanations of the observed nuclear modification pattern. The wide
range of transverse momenta (up to $5.4\,\mathrm{GeV}/c$) of the
present measurement also permits a complementary measurement of the
$p_T$-broadening effect obtained at lower energies.

The paper is divided into four main sections: an overview of the
apparatus, trigger and data samples (Sect.~\ref{sec:data}), a
description of the methods used for the selection and counting of
$J/\psi$s (Sect.~\ref{sec:analysis}), the measurements of the
kinematic distributions (Sect.~\ref{sec:kine}) and the measurements of
the nuclear dependence (Sect.~\ref{sec:adep}), followed by concluding
remarks (Sect.~\ref{sec:conclusions}).

\section{Apparatus, data taking and Monte Carlo simulation} \label{sec:data}
\begin{figure*}[tb]
   \begin{center}
   \includegraphics[width=0.8\textwidth]{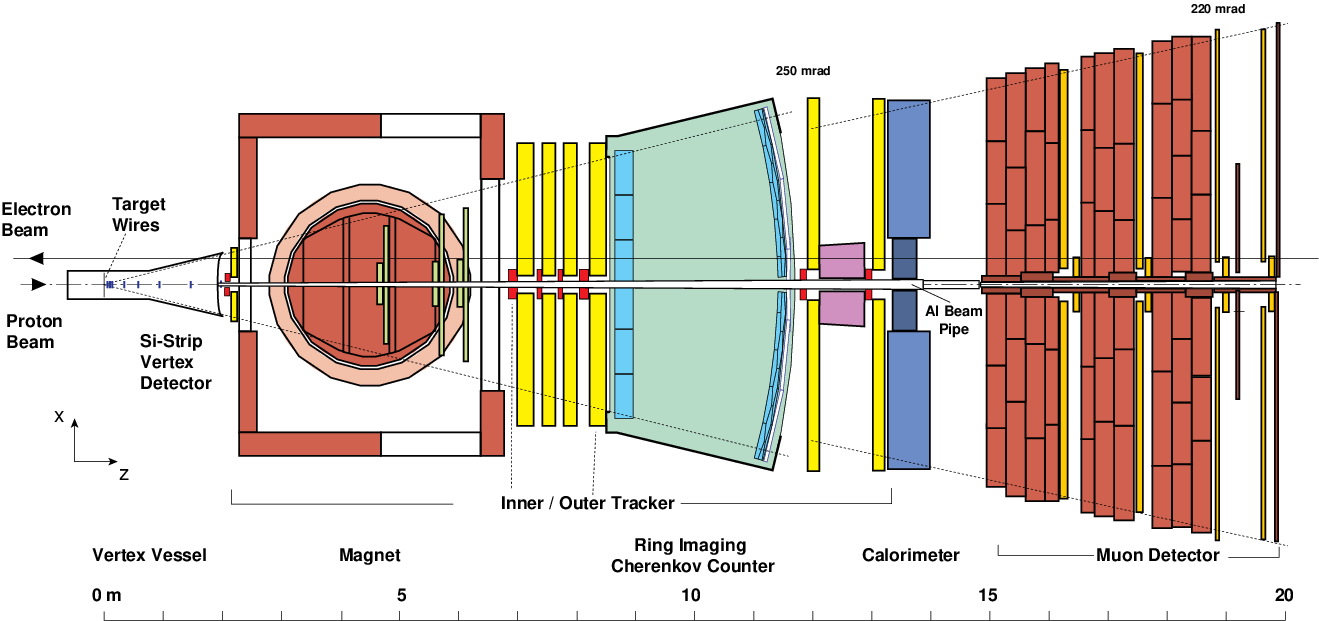}
   \caption{Top view of the HERA-B detector.} \label{fig:detector}
   \end{center}
\end{figure*}

HERA-B was a fixed-target experiment which studied particles produced
in interactions of $920\,\mathrm{GeV/c}$ protons with wire targets positioned
in the halo of the proton beam. The apparatus~\cite{hb_tdr}, shown in
Fig.~\ref{fig:detector}, was a forward spectrometer with
acceptance ranging from $15$ to $220$\,mrad and from $15$ to
$160$\,mrad in the bending ($xz$) and vertical ($yz$)
planes, respectively. Because of this large acceptance and the fact that the $J/\psi$
decay tracks were measured before the muon detector (MUON) and the
electromagnetic calorimeter (ECAL), HERA-B was the first fixed-target
experiment with significant coverage in the region of negative $x_F$
with an accessible range of $x_F \in [-0.34,0.14]$.

The target system~\cite{hb_tar} consisted of eight wires which were
grouped into two stations separated by $4$\,cm along the beam line.
Each wire could be individually steered in the beam halo by a servo
system in order to maintain a constant interaction rate.  A total of
five wires, differing in shape (round or rectangular), dimensions
(between $50\,\mu\mathrm{m}$ and $500\,\mu\mathrm{m}$) and material
(C, Ti and W) were used. Depending on running conditions and run-plan,
either a single wire or a pair of wires was active for any
given data-taking run.  The interaction rate was maintained in the range
of $2$ to $6$\,MHz, depending on beam conditions and target.

The vertex detector (VDS)~\cite{hb_vdet} comprised eight planar
stations of double-sided silicon micro-strip modules, seven of which
were mounted in Roman pots built into the vacuum vessel and operated
at a minimum distance of $10$\,mm from the beam. A track traversed
typically three stations, yielding twelve measurement points in four
stereo views. Vertex resolutions of $450$ and $50$\,$\mu$m in the beam
direction and in the transverse plane, respectively, were
achieved. The eighth station was on a fixed mount immediately
following the exit window of the main vertex system, 2\,m
downstream from the target.

The momenta of charged tracks were measured from their bending
through a vertical magnetic field of integral $2.13$\,Tm. The main
tracker was located between $2$ to $13$\,m downstream of the
target, with one station preceding the magnet, four stations
immediately after the magnet (Pattern Chambers, PCs) and another two
(Trigger Chambers, TCs) after the Ring Imaging Cherenkov detector
(RICH). Each station contained an inner part~\cite{hb_msgc}
(made of micro-strip gas chambers and covering angles less than
$20$\,mrad) which was not included in the trigger system and therefore
does not play a role in the analysis presented here. The outer
part (OTR)~\cite{hb_otr} was composed of honeycomb drift chambers,
with wire pitches of $5$\,mm closer to the beam pipe and $10$\,mm
elsewhere. The final momentum resolution for muons was found to
be $\sigma_{p}/p[\%] = (1.61 \pm 0.02) + (0.0051 \pm 0.0006) 
\cdot p~[\mathrm{GeV}/c]$~\cite{hb_otr}.

The identification of the $J/\psi$ in its dilepton decay modes as well
as the first stage of the trigger system relied mainly on the signals
provided by the ECAL~\cite{ecal} and MUON~\cite{muon} systems.  The
ECAL was a sampling calorimeter using shashlik technology with Pb and
W absorbers sandwiched between scintillator layers. It was divided
into three sections (Inner, Middle and Outer) with cell widths of
$2.2$, $5.5$ and $11$\,cm, respectively, to roughly equalize
occupancies. The inner section used W absorbers, while the middle and
outer sections used Pb. The design was optimized for good
electron/photon energy resolution and for electron-hadron
discrimination. The final energy resolution reached by the detector
can be written in the form $\sigma_E/E = A/\sqrt{E} \oplus B$ ($E$
measured in GeV), with $A = 0.206$, $0.118$ and $0.108$ and $B =
0.012$, $0.014$ and $0.014$, for the Inner, Middle and Outer sections,
respectively. The spatial resolution ranged from $1$ to $10$\,mm
depending on the calorimeter section and on the energy of the
particle~\cite{ecal}.

The MUON system consisted of four tracking stations interleaved with
iron or concrete absorbers. As in the main tracker, two different
technologies were used: gas pixel chambers in the innermost region and
conventional tube chambers in the outer part.  For the last two
stations, not only the anode wires of the tubes but also the segmented
cathodes were read out. Signals from the cathode pads 
were also given as inputs to the trigger.

The RICH~\cite{rich} detector relied on a C$_4$F$_{10}$ radiator and
was used extensively in other analyses for $\pi$/$K$/$p$
separation. It also played a small role in the present
analysis as a means to reject backgrounds in the dilepton analysis,
particularly from kaon decays.

The trigger system selected both $e^+e^-$ and $\mu^+ \mu^-$
signatures and was organized into three levels:
a pretrigger~\cite{ecal,tr_muon}, a First Level Trigger
(FLT~\cite{hb_flt}) and a software-based Second Level Trigger
(SLT~\cite{hb_slt}). The pretrigger used signals from the ECAL and
MUON detectors and required the presence of at least two reconstructed
ECAL clusters with transverse energy above $1.1$\,GeV or
the presence of two muon candidates, defined as
coincidences of projective pads in the last two MUON layers.  Starting
from pretrigger seeds, the FLT attempted to find tracks in a subset of
the OTR tracking layers and required that at least one of the seeds
resulted in a reconstructed track. Starting again from the pretrigger
seeds, the SLT searched for tracks inside regions-of-interest
generated by the pretriggers using all OTR layers and continued the
tracking through the VDS. Finally, at least two fully reconstructed
tracks consistent with the hypothesis of a common vertex were
required. Events passing the SLT were transferred to a computer
farm which provided full online reconstruction of a fraction of the
events for data quality monitoring. The global trigger suppression
factor, $5 \times 10^4$, resulted in an event archival rate of about
$100$\,Hz.

A total of 160 million dilepton triggered events were recorded between
October 2002 and February 2003, together with an approximately fixed
10\,Hz rate of minimum bias events which were used for monitoring and
luminosity determination. The event samples were distributed between
three target materials: carbon ($64\%$), tungsten ($32\%$) and a
small fraction with titanium ($4\%$).

A full Monte Carlo (MC) simulation is used to determine the triggering
(except for FLT, see below), reconstruction and selection
efficiencies. In view of the range of physics topics addressed by the
experiment ($pA$ inelastic interactions, meson decays and heavy flavor
physics), the MC generator is built as a combination of two standard
tools: {\sc Pythia 5.7}~\cite{pythia} for heavy flavor ($b$ or $c$)
quark production in $pN$ interactions and subsequent hadronization and
{\sc Fritiof 7.02}~\cite{fritiof} for light quark production,
secondary interactions in detector materials and generic $pA$
inelastic interactions. The production of the $J/\psi$ is simulated by
first generating the basic hard process $pN \to c \bar c X$ and
subsequent $c \bar c$ hadronization with {\sc Pythia} and then giving
the remaining energy and momentum ($X$) of the interaction to {\sc
Fritiof} for generation of further interactions inside the hit
nucleus.  The generated particles are then given as input to the {\sc
Geant 3.21} based package~\cite{geant} for full simulation of active
(instrumented) and inactive (support structure) elements of the
detector and for the digitization of the electronic signals.

To describe the kinematic characteristics of the produced
$J/\psi$s as accurately as possible, an $x_F$, $p_T$, and decay-angle dependent
weight is assigned to each event and used in the subsequent analysis to force
the simulated $J/\psi$ production and decay distributions to agree
with the corresponding measured distributions for each target
material. The weights are determined by an iterative procedure in
which computed corrections are based on comparisons of MC event
distributions after reconstruction and selection cuts with the
corresponding distributions from data.

The FLT efficiency is derived from an efficiency map which is
determined from the data itself. Since the SLT result is completely
independent of the FLT, and since the FLT triggered on only one of the
two lepton tracks, the efficiency map can be determined by a so-called
tag-and-probe method. Using the efficiency map, each event is assigned a weight
which multiplies the kinematic weight discussed above.

To accurately reproduce the actual working configuration at the time
of data-taking and properly account for time variations of working
conditions, the full data taking period is divided into five
calibration periods of similar lengths, each matched by a
corresponding simulation sample for which the efficiencies of the
individual detector cells are evaluated and given as input to the
MC. The MC samples are reconstructed and analyzed with the same
methods and software packages used for the analysis of the real data.

\section{\boldmath $J/\psi$ selection and counting} \label{sec:analysis}
The electron and muon candidates are selected with common track- and vertex-selection 
criteria while channel-specific methods are applied for lepton identification and the 
treatment of the $J/\psi$ signal.

All tracks passing the SLT are initially considered as lepton
candidates. The track reconstruction procedure consists of finding
straight segments in the VDS and PCs independently, matching them to
each other and also to segments in the TCs. A full, iterative fit of
found tracks is then performed. To reject incorrectly reconstructed or
ghost tracks, loose cuts on the minimum number of hits in the VDS and
OTR, as well as on the $\chi^2 $ probability of the track fit are
applied by the event reconstruction algorithms. The detector
acceptance and trigger requirements effectively limit the momentum and
transverse momentum ranges of lepton tracks to $5 < p <
200\,\mathrm{GeV/c}$ and $0.7 < p_T < 5.0\,\mathrm{GeV/c}$,
respectively. For each event, among all possible pairs of oppositely
charged lepton candidates consistent with a common vertex ($\chi^{2}$
probability greater than $1\%$), only the pair with the best particle
identification (see below) for both leptons is accepted.

\subsection{Dimuon channel}

\begin{figure}
  \includegraphics[width=0.48\textwidth]{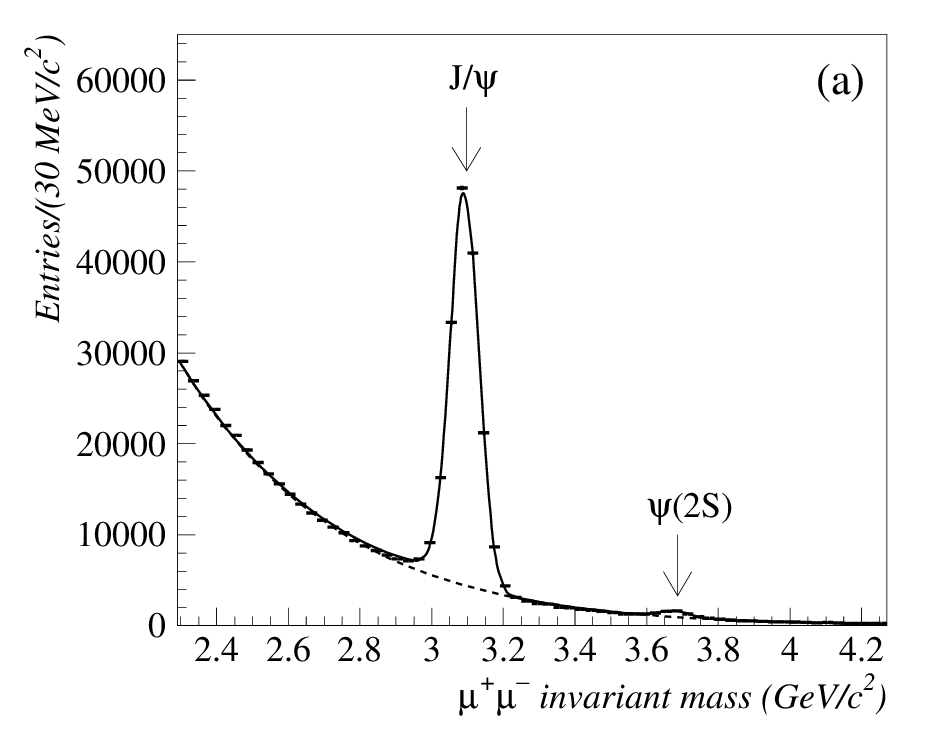}
  \includegraphics[width=0.48\textwidth]{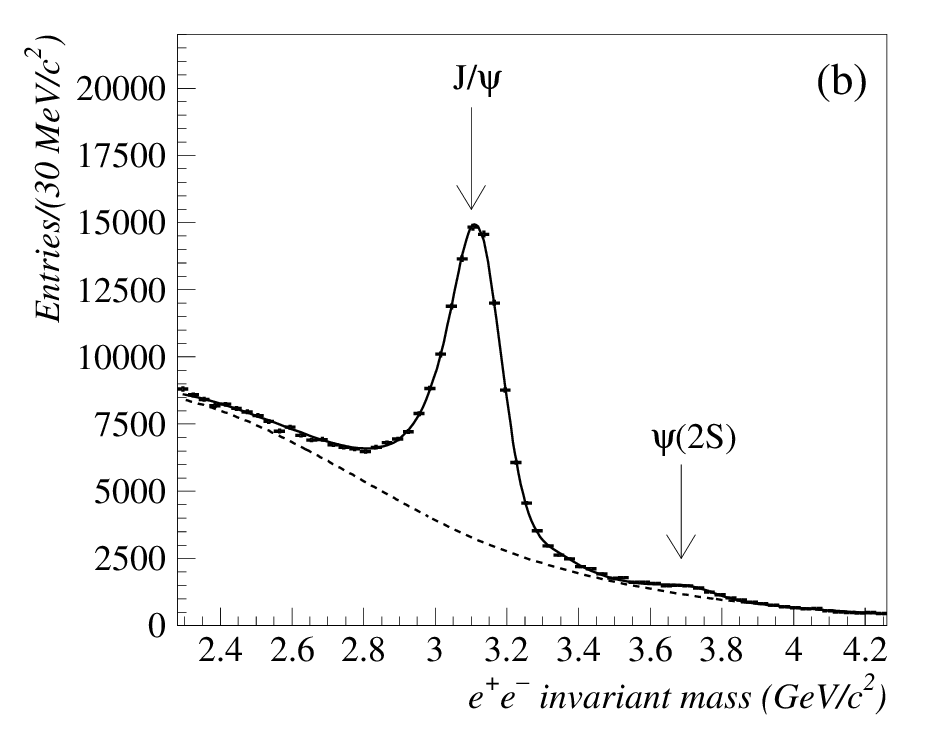}
  \caption{The $\mu^+\mu^-$ (a) and $e^+e^-$ (b) invariant mass
    distributions for the full data sample in the $J/\psi$ mass region. 
    The continuous lines represent the result of fits performed with the
    functions described in the text; the dashed lines are the fitted 
    backgrounds.} 
  \label{invmass}
\end{figure}

Since muons are the only particles having a significant probability
of penetrating through the absorbers of the MUON detector, only minimal
selection cuts are needed to obtain a clean sample (see also~\cite{HBpsiprime}). 
The background of muons from pion and kaon decays is reduced by imposing tighter
cuts on the quality of the track fit and on the matching of track segments
in the VDS, OTR and MUON. Doing so rejects the typical ``broken
trajectories'' produced by decays into leptons. Contamination from kaons is 
further reduced by discarding tracks with high values of the corresponding 
RICH likelihood. After all selection cuts, the background under the $J/\psi$ 
signal is reduced by a factor of $2.5$ with respect to the triggered data 
with a loss of about $11\%$ of the signal.

Fig.~\ref{invmass}(a) shows the resulting dimuon mass spectrum
together with the result of a fit to a sum of three
functions~\cite{spiri}, which model the $J/\psi$ and $\psi(2S)$
signals and the exponential background. The $J/\psi$ and $\psi(2S)$
signals are each modeled as a superposition of three Gaussians with a
common mean which takes into account track resolution and effects of
Moli\`{e}re scattering and a function representing the radiative tail
due to the decay $J/\psi \to \mu^+\mu^-\gamma$~\cite{spiri}. The background 
is described by an exponential of a second-order polynomial. The fitted
position and width of the $J/\psi$ peak are $3.0930\pm
0.0002\,{\rm GeV}/{\rm c}^2$ and $40 \pm 1\,\mathrm{MeV/c^2}$, respectively.

\begin{table}
\begin{center}
    \begin{tabular}{crrrr}
      Channel&   C      &    Ti   &    W    & Total  \\
      \hline
      $\mu^+ \mu^-$ &  94800    &  8060   &  48100  & 152000 \\
      $e^+ e^-$     &  57700    &  4280   &  25300  & 87200  \\ \hline

    \end{tabular}

  \caption{The numbers of reconstructed $J/\psi$s after all selection cuts
    in the dimuon and dielectron channels, and for different target
    materials.} \label{tb:njpsi}
\end{center}
\end{table}

\begin{figure*}[htb]
  \begin{center}
  \includegraphics[width=0.75\textwidth]{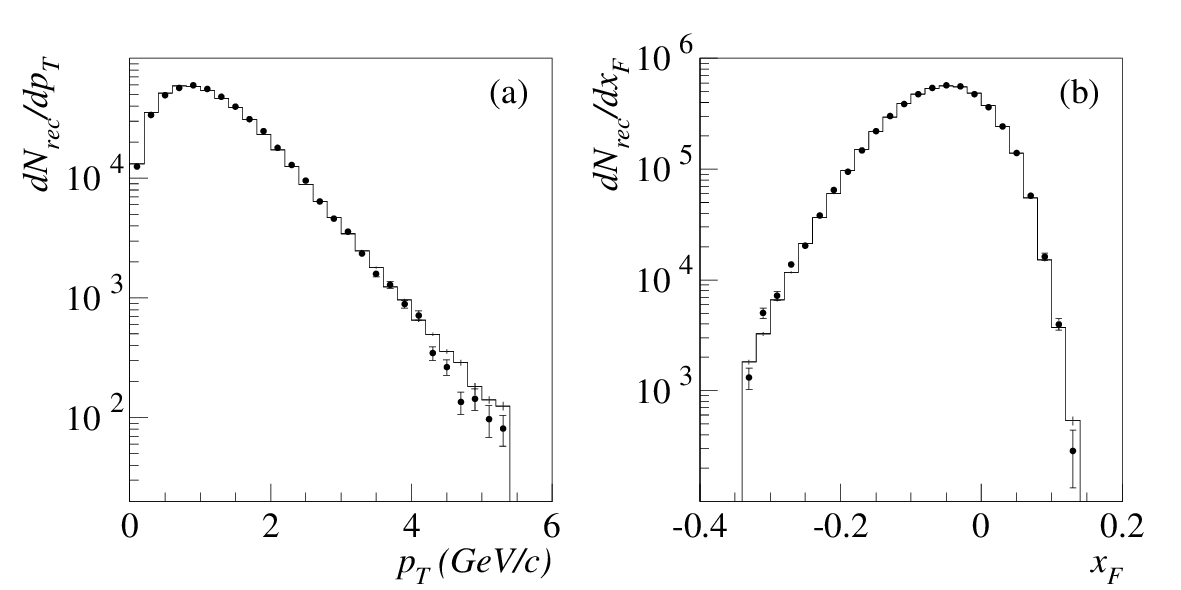}
  \includegraphics[width=0.75\textwidth]{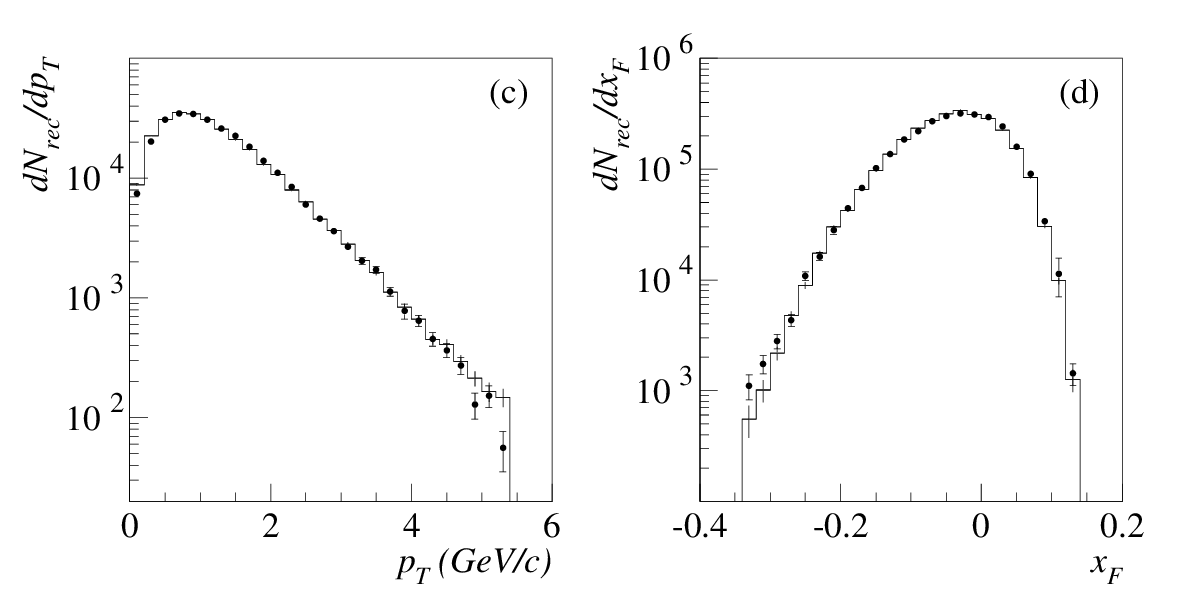}
  \caption{Distribution of the number of $J/\psi$s reconstructed in
    the carbon target sample (points) and in the corresponding MC data
    (histograms, arbitrarily re-normalized) as a function of $p_T$ and
    $x_F$. (a) and (b): muon channel, (c) and (d): electron channel.}
  \label{fig:accep}
  \end{center}
\end{figure*}

\subsection{Dielectron channel}

The $J/\psi$ selection in the dielectron channel is affected by major
background contributions from charged hadrons which produce energetic
ECAL clusters and by overlapping photon and charged-hadron energy
deposits in the ECAL. For this reason, the electron identification
requirements were the subject of careful optimization studies which
resulted in much more stringent selection cuts than for the muon sample.

A cut on the transverse energy of the ECAL cluster ($E_T> 1.15$\,GeV)
is applied in order to mask different threshold cuts applied at
the pretrigger level for the various acquisition periods.

The reconstructed momentum vectors of electrons and positrons are
corrected for energy loss from brems\-strahlung (BR) emission in the
materials in front of the magnet. For each electron track, an attempt
to identify an ECAL cluster due to a BR photon is made by looking for
an energy deposition in coincidence with the extrapolation of the
associated VDS track segment to the ECAL. Any recovered energy (about
$18\%$ of the initial electron energy on average) is then added to the
momentum measured by the tracking system. Since BR is a clear
signature for an electron track, it is also exploited to obtain
substantial background reduction which is essential for accurate
$J/\psi$ counting. The baseline results of the analysis are obtained
by requiring that at least one lepton of a decaying $J/\psi$ has an
associated BR cluster. This requirement reduces the signal by about
$30\%$ and suppresses the background by more than a factor of
two. Alternative requirements (no BR requirement, only one, or both
electrons emitting BR) lead to very different background shapes and
amounts. The differences are exploited for systematic studies on the
stability of signal counting and on the correctness of the MC
simulation.

Adjustments to the measured momenta of electron tracks were applied to
compensate for differences in multiple scattering between electrons
and muons since the track-fitter had been calibrated for muons. For
this purpose, a correction map, determined from the shift of the
$J/\psi$ peak position in different kinematic regions was used.

Additional selection cuts are applied to further improve the
significance of the dielectron signal. A particularly discriminating
variable is the ratio $E/p$, where $E$ is the energy of an ECAL
cluster and $p$ is the momentum of the associated track. The $E/p$
distribution for electrons has a Gaussian shape with a mean value
close to 1 and width varying between $6.4\%$ and $7.4\%$ depending on
calorimeter sector.  Values of $E/p$ much lower than $1$ are mainly
due to particles, mostly hadrons, which release only part of their
energy in the calorimeter.  Further selection variables used in the
analysis are the distances $\Delta x$ and $\Delta y$ -- along the $x$
and $y$ directions -- between the reconstructed cluster and the track
position extrapolated to the ECAL. The $\Delta x$ and $\Delta y$
distributions for electrons are, apart from a small tail,
well described by Gaussians centered at zero with widths between
$0.2$ and $1.0$\,cm depending on calorimeter sector. Cuts on these
quantities lead to a significant reduction of the contamination from
hadrons and random cluster-track matches which are characterized by
significantly wider distributions. The selection of the candidate
electron-positron pairs is further refined by putting an upper bound
on the distance of closest approach ($\Delta b$) between the two
accepted tracks near the vertex.

All the requirements described above have been simultaneously
optimized by maximizing the significance $S/\sqrt{S+B}$ of the
$J/\psi$ signal ($S$) -- taken from the MC (scaled to the number of 
$J/\psi$ in the data) -- with respect to the background ($B$) -- evaluated 
from the data. The optimal ranges for the different cut variables depend 
on the number (one or two) of BR clusters associated to the 
electron-positron pair. When both electron and positron have a BR cluster
correlated to the track, where the cluster position was determined
by hierarchical clustering~\cite{ecal}, the event is already rather 
cleanly reconstructed and only one additional request (for each lepton candidate), 
$(E/p - 1)/\sigma_{E/p} > -3.6$, is applied. When only one of the two possible 
BR clusters is found, the accepted ranges are determined for each lepton as 
$-3.6  < (E/p - 1)/\sigma_{E/p} < 5.4$, $|\Delta x|/\sigma_{\Delta x} < 7.0$, 
$|\Delta y|/\sigma_{\Delta y} < 3.3$ and $\Delta b < 370\,\mu {\rm m}$, respectively.

The combined selection cuts increase the $S/B$ ratio of the $J/\psi$
by about a factor of $10$ with respect to triggered events, and have
an overall efficiency of $(45 \pm 4)\%$ as evaluated using the data and
verified with the simulation -- the rather large uncertainty is due to
the difficulty of counting the signal when no cuts are applied. As can
be seen in Fig.~\ref{invmass}(b), the significance of the optimized
$J/\psi$ signal, although less than that of the muon channel shown in
Fig.~\ref{invmass}(a), is nonetheless such that the electron sample
significantly enhances the statistical precision of the final results. The
method adopted for counting the signal uses a Gaussian shape for the
right part of the peak and a Breit-Wigner shape for the left part to
account for the sizable asymmetry of the signal caused by
missing BR energy and the contribution of the radiative decay $J/\psi
\to e^+e^- \gamma$. The background is parametrized with a Gaussian at
lower mass values and an exponential at higher mass, joined together
such that the resulting curve is smooth. The position and width of the
$J/\psi$ peak as determined from the fit are $3.110 \pm 0.001$ ${\rm
GeV}/{\rm c}^2$ and $72 \pm 1\,\mathrm{MeV/c^2}$, respectively.

\bigskip

The yields  of selected  $J/\psi$ candidates  for the two decay channels and for 
each target material are listed in Table~\ref{tb:njpsi}. Fig.~\ref{fig:accep} shows 
(for the carbon data) a comparison  between data and MC of the distributions of 
reconstructed $J/\psi$s as a function of the kinematic variables $p_T$ and $x_F$.

\section{Kinematic distributions} \label{sec:kine}
\subsection{Results}
\label{sec:kine_results}

\begin{figure}
  \begin{center}
  \includegraphics[width=0.46\textwidth]{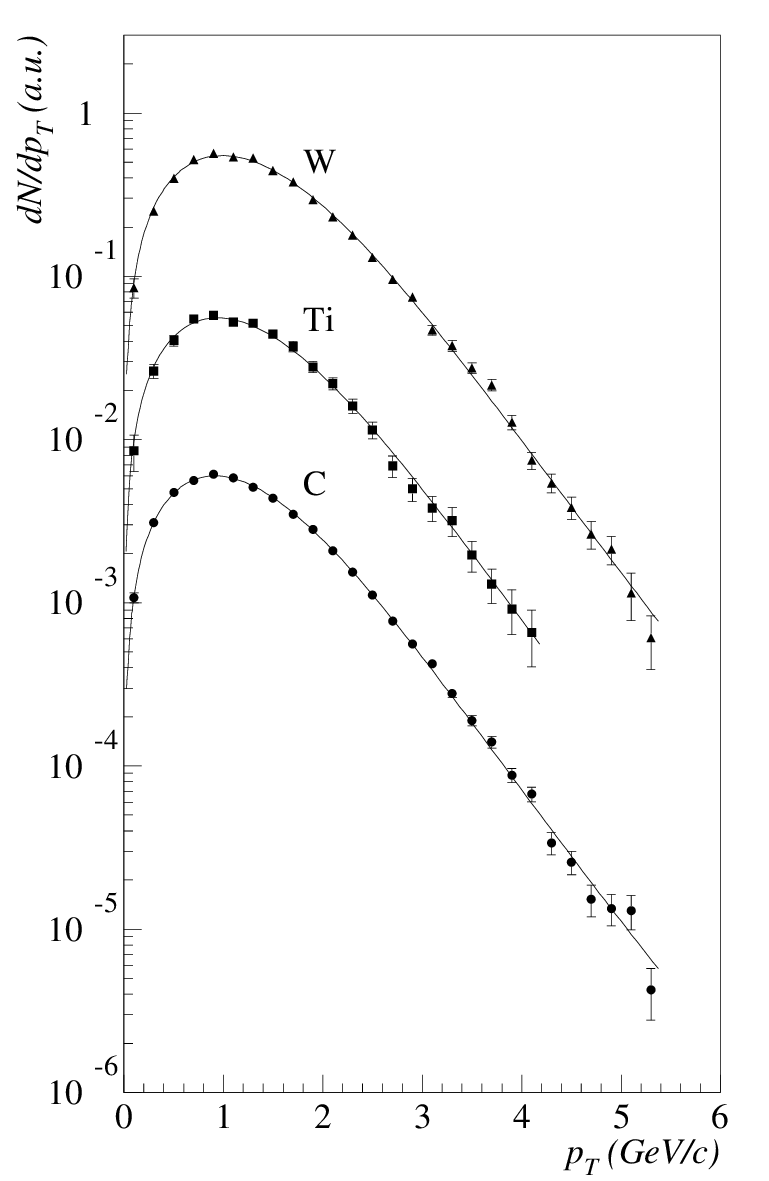}
  \caption{ Inclusive $p_T$ distributions of $J/\psi$ mesons for three target
    materials with arbitrary normalizations. 
    The error bars represent the combination of statistical 
    and systematic uncertainties. 
    The interpolating lines are the results of a
    simultaneous fit of the three $p_T$ distributions to
    Eq.~\ref{eq:ptfit} performed with the method described in
    Sec.~\protect{\ref{sec:kindist}.}} 
  \label{fig:pt_ctiw}
  \end{center}
\end{figure}

\begin{figure}
  \begin{center}
  \includegraphics[width=0.48\textwidth]{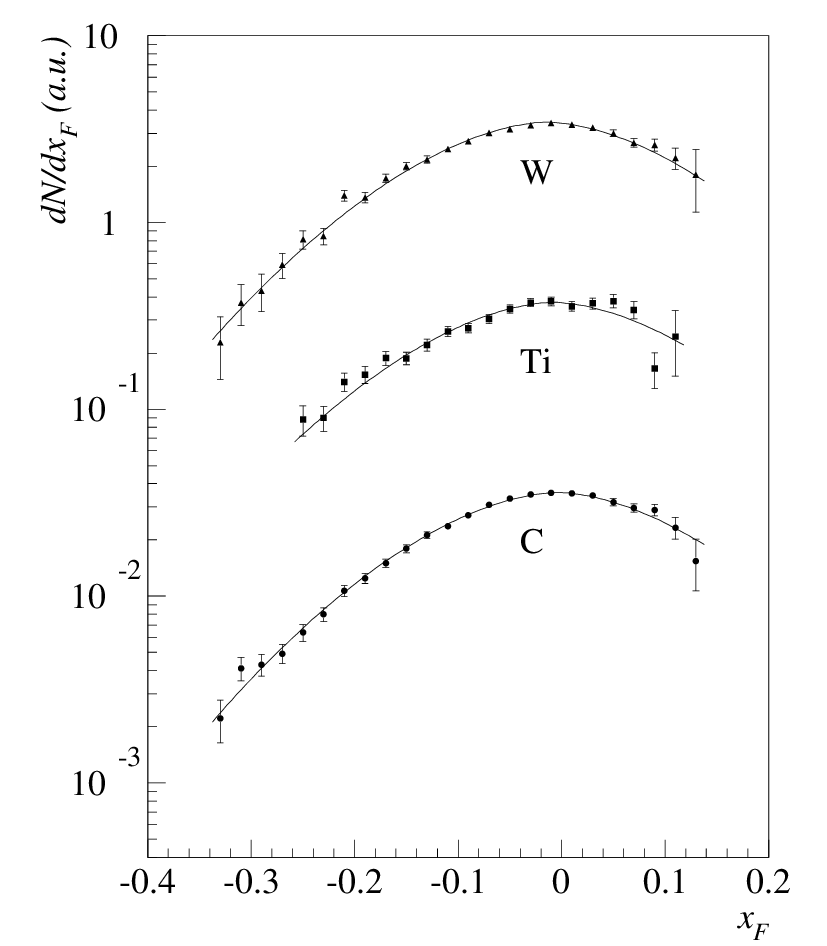}
  \caption{Inclusive $x_F$ distributions of $J/\psi$ mesons for the three target
    materials with arbitrary normalizations. The error bars
    represent the combination of statistical and systematic
    uncertainties. The interpolating lines are the results of a
    simultaneous fit of the three $x_F$ distributions to 
    Eq.~\ref{eq:xffit} performed with the method described in
    Sec.~\ref{sec:kindist}.} 
  \label{fig:xf_ctiw}
  \end{center}
\end{figure}

\begin{table*}
    \begin{center}
    \begin{tabular}{cc |
                       @{\hspace{7mm}}r@{.}l  @{\hspace{1mm}$\pm$\hspace{1mm}}r@{.}l @{\hspace{1mm}$\pm$\hspace{1mm}}r@{.}l    
                       @{\hspace{7mm}}r@{.}l  @{\hspace{1mm}$\pm$\hspace{1mm}}r@{.}l @{\hspace{1mm}$\pm$\hspace{1mm}}r@{.}l
                       @{\hspace{7mm}}r@{.}l  @{\hspace{1mm}$\pm$\hspace{1mm}}r@{.}l @{\hspace{1mm}$\pm$\hspace{1mm}}r@{.}l}
       \multicolumn{2}{c|@{\hspace{7mm}}}{$p_T$ (GeV$/c$)} 
     & \multicolumn{6}{c}{C $(\times 10^{-2})$}  
     & \multicolumn{6}{c}{ Ti  $(\times 10^{-2})$} 
     & \multicolumn{6}{c}{ W  $(\times 10^{-2})$}                 \\
       min & 
       \multicolumn{1}{c|@{\hspace{7mm}}}{max} & \multicolumn{18}{c}{}                          \\
      \hline
      0.0&0.2  & 10&74 & 0&28 & 0&72  &   8&5 & 0&9 & 1&9    &  8&5  & 0&3  & 1&1 \\
      0.2&0.4  & 30&94 & 0&46 & 0&69  &  26&3 & 1&7 & 2&0    & 25&1  & 0&7  & 1&1 \\
      0.4&0.6  & 47&36 & 0&61 & 0&66  &  40&4 & 2&1 & 2&1    & 39&9  & 0&7  & 1&1 \\
      0.6&0.8  & 55&90 & 0&73 & 0&62  &  54&8 & 2&3 & 2&1    & 51&9  & 0&9  & 1&1 \\
      0.8&1.0  & 61&10 & 0&66 & 0&58  &  57&6 & 2&3 & 2&1    & 56&5  & 1&0  & 1&1 \\
      1.0&1.2  & 58&16 & 0&70 & 0&53  &  52&3 & 2&1 & 2&0    & 53&8  & 1&0  & 1&0 \\
      1.2&1.4  & 51&11 & 0&66 & 0&49  &  51&5 & 2&2 & 1&9    & 52&8  & 1&0  & 1&0 \\
      1.4&1.6  & 43&57 & 0&62 & 0&44  &  44&2 & 1&9 & 1&8    & 44&58 & 0&85 & 0&87 \\
      1.6&1.8  & 34&84 & 0&58 & 0&40  &  37&1 & 2&0 & 1&6    & 37&89 & 0&76 & 0&77 \\
      1.8&2.0  & 28&17 & 0&53 & 0&35  &  27&9 & 1&5 & 1&4    & 29&46 & 0&66 & 0&68 \\
      2.0&2.2  & 20&78 & 0&42 & 0&31  &  22&0 & 1&4 & 1&3    & 23&02 & 0&66 & 0&58 \\
      2.2&2.4  & 15&37 & 0&33 & 0&27  &  16&0 & 1&2 & 1&1    & 17&81 & 0&52 & 0&49 \\
      2.4&2.6  & 11&18 & 0&26 & 0&23  &  11&4 & 1&0 & 0&9    & 13&06 & 0&44 & 0&40 \\
      2.6&2.8  &  7&73 & 0&23 & 0&20  &  6&89 & 0&72& 0&71   &  9&56 & 0&37 & 0&33 \\
      2.8&3.0  &  5&59 & 0&20 & 0&17  &  4&99 & 0&57& 0&57   &  7&45 & 0&36 & 0&26 \\
      3.0&3.2  &  4&22 & 0&16 & 0&14  &  3&80 & 0&50& 0&44   &  4&68 & 0&24 & 0&20 \\
      3.2&3.4  &  2&79 & 0&12 & 0&12  &  3&18 & 0&56& 0&33   &  3&77 & 0&25 & 0&15 \\
      3.4&3.6  &  1&90 & 0&10 & 0&09  &  1&96 & 0&34& 0&25   &  2&75 & 0&18 & 0&11 \\
      3.6&3.8  &  1&402& 0&082& 0&077 &  1&30 & 0&25& 0&18   &  2&16 & 0&15 & 0&08 \\
      3.8&4.0  &  0&879& 0&064& 0&062 &  0&92 & 0&25& 0&13   &  1&27 & 0&11 & 0&06 \\
      4.0&4.2  &  0&672& 0&051& 0&049 &  0&65 & 0&23& 0&09   &  0&746& 0&081& 0&042 \\
      4.2&4.4  &  0&338& 0&036& 0&039 &  \multicolumn{6}{c}{}&  0&542& 0&067& 0&029 \\
      4.4&4.6  &  0&257& 0&030& 0&030 &  \multicolumn{6}{c}{}&  0&383& 0&056& 0&019 \\
      4.6&4.8  &  0&153& 0&025& 0&023 &  \multicolumn{6}{c}{}&  0&263& 0&049& 0&013 \\
      4.8&5.0  &  0&134& 0&023& 0&018 &  \multicolumn{6}{c}{}&  0&212& 0&041& 0&008 \\
      5.0&5.2  &  0&130& 0&028& 0&013 &  \multicolumn{6}{c}{}&  0&115& 0&036& 0&005 \\
      5.2&5.4  &  0&042& 0&011& 0&010 &  \multicolumn{6}{c}{}&  0&061& 0&022& 0&003 \\ \hline
    \end{tabular}
  \caption{$J/\psi$ $p_T$ distributions ($dN/dp_T$, normalized to their
    integrals over the measured range) for three target materials with statistical and
    systematic uncertainties.} 
  \label{tb:pt_ctiw}
  \end{center}
\end{table*}

\begin{table*}
    \begin{center}
    \begin{tabular}{
                    r@{.}l 
                    r@{.}l|
                    @{\hspace{7mm}}r@{.}l  @{\hspace{1mm}$\pm$\hspace{1mm}}r@{.}l @{\hspace{1mm}$\pm$\hspace{1mm}}r@{.}l    
                    @{\hspace{7mm}}r@{.}l  @{\hspace{1mm}$\pm$\hspace{1mm}}r@{.}l @{\hspace{1mm}$\pm$\hspace{1mm}}r@{.}l
                    @{\hspace{7mm}}r@{.}l  @{\hspace{1mm}$\pm$\hspace{1mm}}r@{.}l @{\hspace{1mm}$\pm$\hspace{1mm}}r@{.}l}
       \multicolumn{4}{c|@{\hspace{7mm}}}{$x_F$} 
     & \multicolumn{6}{c}{ C }  
     & \multicolumn{6}{c}{ Ti } 
     & \multicolumn{6}{c}{ W  }                 \\
       \multicolumn{2}{c}{min }
     & \multicolumn{2}{c|@{\hspace{7mm}}}{max } 
     & \multicolumn{18}{c}{}                     \\
      \hline
       -0&34 & -0&32 & 0&221 & 0&045 & 0&035 &\multicolumn{6}{c}{}& 0&228 & 0&061 & 0&058 \\
       -0&32 & -0&30 & 0&412 & 0&044 & 0&039 &\multicolumn{6}{c}{}& 0&372 & 0&070 & 0&059 \\
       -0&30 & -0&28 & 0&429 & 0&037 & 0&044 &\multicolumn{6}{c}{}& 0&431 & 0&076 & 0&061 \\
       -0&28 & -0&26 & 0&492 & 0&031 & 0&049 &\multicolumn{6}{c}{}& 0&593 & 0&065 & 0&062 \\
       -0&26 & -0&24 & 0&639 & 0&036 & 0&054 &0&88 & 0&16 & 0&05  & 0&812 & 0&065 & 0&063 \\
       -0&24 & -0&22 & 0&798 & 0&035 & 0&058 &0&90 & 0&13 & 0&05  & 0&846 & 0&057 & 0&065 \\
       -0&22 & -0&20 & 1&067 & 0&036 & 0&063 &1&41 & 0&15 & 0&06  & 1&393 & 0&066 & 0&066 \\
       -0&20 & -0&18 & 1&247 & 0&034 & 0&068 &1&53 & 0&15 & 0&06  & 1&361 & 0&059 & 0&068 \\
       -0&18 & -0&16 & 1&504 & 0&035 & 0&073 &1&88 & 0&14 & 0&07  & 1&725 & 0&057 & 0&069 \\
       -0&16 & -0&14 & 1&791 & 0&030 & 0&078 &1&87 & 0&12 & 0&07  & 2&000 & 0&063 & 0&070 \\
       -0&14 & -0&12 & 2&119 & 0&033 & 0&082 &2&21 & 0&14 & 0&08  & 2&171 & 0&065 & 0&072 \\
       -0&12 & -0&10 & 2&374 & 0&031 & 0&087 &2&61 & 0&13 & 0&08  & 2&477 & 0&056 & 0&073 \\
       -0&10 & -0&08 & 2&710 & 0&033 & 0&092 &2&71 & 0&13 & 0&09  & 2&713 & 0&056 & 0&074 \\
       -0&08 & -0&06 & 3&074 & 0&039 & 0&097 &3&04 & 0&13 & 0&09  & 3&022 & 0&051 & 0&076 \\
       -0&06 & -0&04 & 3&33  & 0&05  & 0&10  &3&45 & 0&14 & 0&10  & 3&161 & 0&052 & 0&077 \\
       -0&04 & -0&02 & 3&51  & 0&04  & 0&11  &3&73 & 0&15 & 0&10  & 3&318 & 0&055 & 0&078 \\
       -0&02 &  0&00 & 3&56  & 0&04  & 0&11  &3&78 & 0&16 & 0&11  & 3&413 & 0&057 & 0&080 \\
        0&00 &  0&02 & 3&54  & 0&05  & 0&12  &3&56 & 0&18 & 0&11  & 3&349 & 0&065 & 0&081 \\
        0&02 &  0&04 & 3&47  & 0&05  & 0&12  &3&68 & 0&22 & 0&12  & 3&207 & 0&071 & 0&082 \\
        0&04 &  0&06 & 3&18  & 0&07  & 0&13  &3&80 & 0&30 & 0&12  & 2&998 & 0&093 & 0&084 \\
        0&06 &  0&08 & 2&97  & 0&08  & 0&13  &3&40 & 0&33 & 0&13  & 2&67  & 0&11  & 0&09 \\
        0&08 &  0&10 & 2&90  & 0&14  & 0&13  &1&65 & 0&33 & 0&13  & 2&60  & 0&17  & 0&09 \\
        0&10 &  0&12 & 2&33  & 0&28  & 0&14  &2&44 & 0&93 & 0&14  & 2&21  & 0&28  & 0&09 \\
        0&12 &  0&14 & 1&54  & 0&45  & 0&14  &\multicolumn{6}{c}{}& 1&80  & 0&65  & 0&09 \\ \hline
    \end{tabular}
  \caption{$J/\psi$ $x_F$ distributions ($dN/dx_F$, normalized to their
    integrals over the measured range) for the three target materials with statistical and
    systematic uncertainties.} 
  \label{tb:xf_ctiw}
  \end{center}
\end{table*}

The present analysis adopts the degrees of freedom $p_T$, $x_F$ and
$\Phi$ (azimuthal production angle) for the description of the $J/\psi$
production kinematics. Single-variable distributions are obtained
according to the formula (here written e.g. for $x_F$)
\begin{equation}
  \frac{dN_{J/\psi}}{d x_F}(x_F) = \frac{ \Delta
    N^{\mathrm{rec}}_{J/\psi}(x_F)} { \epsilon_{J/\psi}(x_F) \
    \Delta x_F },  \label{eq:normdistr}
\end{equation}
where $\Delta N^{\mathrm{rec}}_{J/\psi}(x_F)$ is the fraction of $J/\psi$s reconstructed in 
a given $x_F$ interval of width $\Delta x_F$ and $\epsilon_{J/\psi}(x_F)$ is the global 
(trigger, reconstruction and selection) efficiency for that interval integrated over all 
other kinematic variables (including the $J/\psi$ decay degrees of freedom) using the tuned MC. 
In each case, the signal is also integrated over all other kinematic variables. All 
distributions are normalized to unit area\footnote{
The absolute $J/\psi$ yield in proton-nucleus collisions at $920$\,GeV$/c$ was the subject 
of a measurement performed using minimum bias data~\cite{jpsi_xsect_HB}.}.

The final $p_T$ and $x_F$ distributions for the three materials are
shown in Figs.~\ref{fig:pt_ctiw} and~\ref{fig:xf_ctiw}. The error bars
include statistical and systematic uncertainties added in
quadrature. The corresponding numbers can be found in
Tables~\ref{tb:pt_ctiw} and~\ref{tb:xf_ctiw}. The final distributions
and systematic uncertainties are evaluated by fixing the input
parameters and assumptions of the analysis to a variety of values within
their range of uncertainties (see list below) and
carrying through the full analysis. The kinematic distributions are
finally obtained by averaging over decay channels for each set of input
parameters and assumptions.  The central value for each bin is the
mid-point of the distribution of values thus obtained and the
systematic uncertainty is the maximum spread of the obtained values
divided by $\sqrt{12}$. The stability tests are described in the
following list.

\begin{itemize}
  \item The impact of \emph{selection and optimization} requirements
  is evaluated by changing the cuts on momentum and transverse
  momentum of muon and electron candidates with respect to the
  intrinsic thresholds of the trigger selection, and by scanning
  systematically the values of all cut variables used for the
  optimization of the dimuon and dielectron signals (including, for
  the latter, different BR requirements).  

\item The uncertainty
  associated to the \emph{signal counting method} has been estimated
  from the variation of the results obtained with the adoption of
  modified background and signal functions. Special attention is given
  to the background evaluation of the dielectron channel: as an
  alternative to the fit with an assumed background function, a
  background shape constructed by mixing real events (combining each
  track with one of opposite-charge from a different event) has been
  used in an unbinned maximum-likelihood fit of the invariant mass
  spectrum. A further cross-check is represented by the comparison
  between the efficiency-corrected $J/\psi$ yield obtained with
  different BR requirements (and therefore different background
  shapes). When the BR requirement is removed, the evaluation of the
  number of $J/\psi$s becomes less stable due to increased background,
  but the variation of the efficiency-corrected yield with respect to
  the standard selection is estimated to be lower than $5\%$.

\item The systematic uncertainties also account for the
  stability of the results when specific \emph{acquisition periods and
  conditions} are selected. A large subsample of the collected events
  was produced on two target wires of different materials operated
  simultaneously. The comparison of these data with those acquired
  with a single target provides an indication of the extent to which the
  measurements are affected by variations of the experimental
  conditions.

\item The results are sensitive to the
  shape of the \emph{$J/\psi$ decay angular distribution} assumed in
  the MC generator. The hypothesis -- made by previous experiments --
  that the $J/\psi$ is produced in an unpolarized state is not
  supported by the HERA-B data~\cite{PFconf}. There is, moreover, an
  indication that the polarization increases in magnitude with
  decreasing $p_T$, while no significant $x_F$ dependence is
  found. Since a longitudinally polarized $J/\psi$ is detected more
  efficiently due to the lower probability that its decay leptons
  escape detection by passing through the uninstrumented region near
  the beam, the kinematic dependence of the polarization assumed in
  the MC influences the shape of the efficiency-corrected $p_T$ and/or
  $x_F$ spectra.  The systematic stability tests therefore include a
  variety of different assumptions for polarization (including
  longitudinal, $p_T$-dependent polarization -- also considering the
  possibility of an $A$-dependent polarization -- and the absence of
  polarization).
\end{itemize}

\begin{figure}
  \begin{center}
  \includegraphics[width=0.48\textwidth]{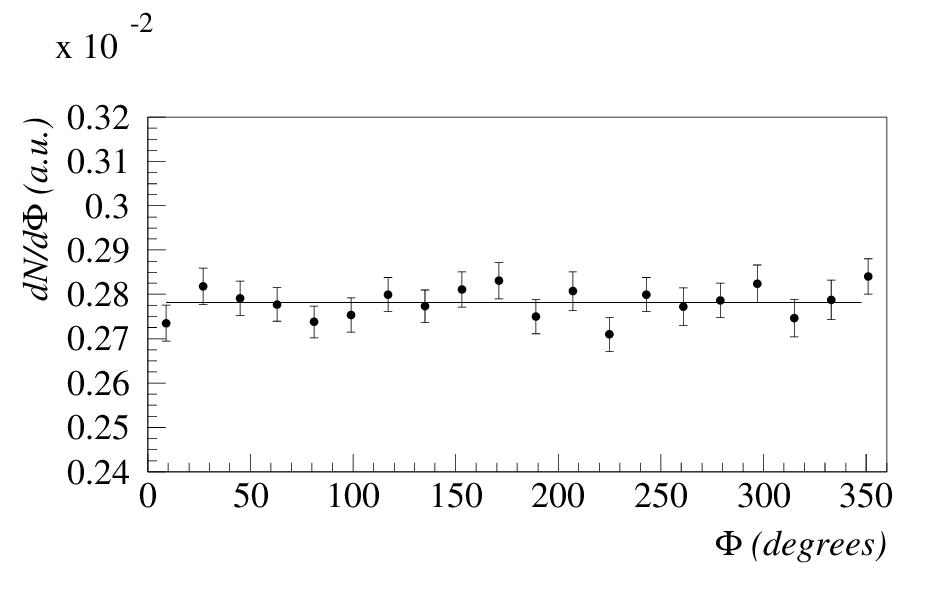}
  \caption{Distribution of the azimuthal production angle of the $J/\psi$s fitted
    to a constant. The error bars include only statistical contributions.}
  \label{fig:Phi}
  \end{center}
\end{figure}

The distribution of the azimuthal production angle $\Phi$ has been evaluated as a systematic 
check of the uniformity of the MC description of the geometrical acceptance. 
Fig.~\ref{fig:Phi} shows the result obtained when combining the full data 
reconstructed in both decay channels: the points are consistent, within the 
statistical uncertainties, with the expected flat distribution.

\begin{table*}
    \begin{center}
    \begin{tabular}{@{\hspace{1mm}}c@{\hspace{2mm}}c|
                    @{\hspace{5mm}}r@{.}l  @{\hspace{1mm}$\pm$\hspace{1mm}}r@{.}l @{\hspace{1mm}$\pm$\hspace{1mm}}r@{.}l    
                    @{\hspace{5mm}}r@{.}l  @{\hspace{1mm}$\pm$\hspace{1mm}}r@{.}l @{\hspace{1mm}$\pm$\hspace{1mm}}r@{.}l
                    @{\hspace{5mm}}r@{.}l  @{\hspace{1mm}$\pm$\hspace{1mm}}r@{.}l @{\hspace{1mm}$\pm$\hspace{1mm}}r@{.}l}
      Channel      & Param.
     & \multicolumn{6}{c}{ C }  
     & \multicolumn{6}{c}{ Ti } 
     & \multicolumn{6}{c}{ W  }                 \\
      \hline
      $\mu^+ \mu^-$&$\langle p_T^2 \rangle $& 2&141 & 0&011 & 0&014  & 2&200 & 0&044 & 0&015  &2&435 & 0&017 & 0&026\\
                   & $\beta$                & 7&31  & 0&20  & 0&14   & 9&3   & 1&6   & 0&5    &8&09  & 0&34 & 0&24    \\
      \hline
      $e^+ e^-$&$\langle p_T^2 \rangle $    &2&149  & 0&019 & 0&025  & 2&220 & 0&069 & 0&044  &2&460 & 0&034 & 0&041\\
               &$\beta$                     &7&14   & 0&29  & 0&13   & 9&0   & 2&2   & 0&3    &8&81  & 0&81  & 0&76 \\
      \hline
      comb.      &$\langle p_T^2 \rangle $  &2&141  &0&009  & 0&015  & 2&204 & 0&036 & 0&018  &2&432 & 0&015 &0&028\\
                 &$\beta$                   &7&28   &0&16   & 0&13   & 9&3   & 1&3   & 0&3    & 8&13 & 0&30  &0&28 \\
      \hline
      \hline
      &$w_{x_F}$                            & 0&1464 & 0&0026 & 0&0023 & 0&1453 & 0&0088 & 0&0051  & 0&1592 & 0&0037 & 0&0018\\
      $\mu^+ \mu^-$&$\Delta x_F$            &-0&0030 & 0&0020 & 0&0017 &-0&0076 & 0&0056 & 0&0021  &-0&0095 & 0&0027 & 0&0010  \\
                   & $\gamma      $         & 1&699  & 0&039  & 0&014  & 1&45   & 0&15   & 0&02    & 1&810  & 0&071  & 0&016\\
      \hline
                   & $w_{x_F}$              & 0&1482 & 0&0042 & 0&0014 & 0&149  & 0&015  & 0&006   & 0&1599 & 0&0081 & 0&0052\\
      $e^+ e^-$    &$\Delta x_F$            &-0&0016 & 0&0029 & 0&0034 &-0&006  & 0&010  & 0&003   &-0&0056 & 0&0058 & 0&0017  \\
                   &$\gamma      $          & 1&749  & 0&075  & 0&031  & 1&54   & 0&30   & 0&02    & 1&80   & 0&17   & 0&11\\
      \hline
                   &$w_{x_F}$               & 0&1468 & 0&0022 & 0&0016 & 0&1482 & 0&0079 & 0&0028  & 0&1588 & 0&0033 & 0&0019\\
      comb.        &$\Delta x_F$            &-0&0024 & 0&0016 & 0&0022 &-0&0052 & 0&0051 & 0&0015  &-0&0096 & 0&0024 & 0&0012 \\
      &$\gamma      $                       & 1&723  & 0&036  & 0&011  & 1&48   & 0&14   & 0&01    & 1&820  & 0&063  & 0&018\\\hline
    \end{tabular}

  \caption{Parameter values obtained from the fit of the kinematic
  distributions of each of the three target samples to the functions
  described in the text (Eq.s~\ref{eq:ptfit} and \ref{eq:xffit}). The
  first of the given uncertainty ranges is statistical and the second
  is systematic.  } \label{tb:distr_params} \end{center}
\end{table*}

\begin{figure}
  \begin{center}
  \includegraphics[width=0.47\textwidth]{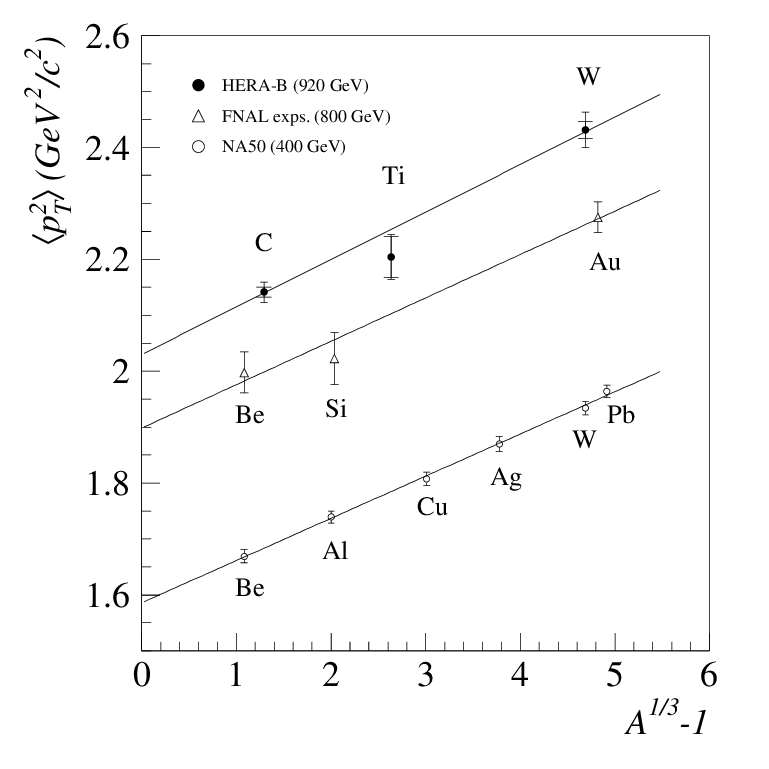}
  \caption{The $\langle p_T^2 \rangle $ of the produced $J/\psi$s as a
    function of $A^{1/3}-1$ (see Eq.~\ref{eq:ptbroad}). The results of the 
    present analysis (black filled circles with total and statistical 
    uncertainties) are compared to previous measurements performed with 
    different beam energies at Fermilab~\cite{FNAL} and at the SPS~\cite{na50}. 
    The data are fitted with linear functions.} 
  \label{fig:pt2_vs_a13}
  \end{center}
\end{figure}

\begin{figure}
  \begin{center}
  \includegraphics[width=0.4\textwidth]{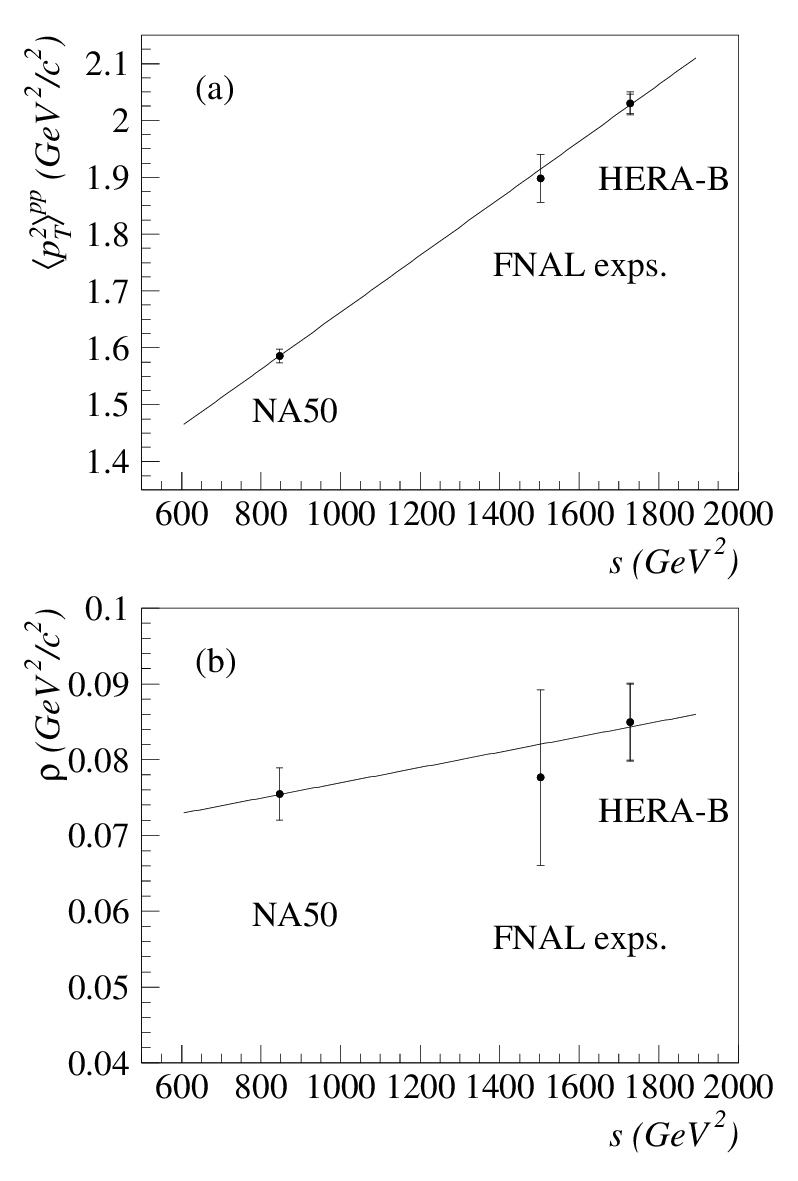}
  \caption{Energy dependence of the parameters $\langle p_T^2 \rangle^{pp}$ (a) 
    and $\rho$ (b) as defined in Eq.~\ref{eq:ptbroad} describing the $p_T$-broadening 
    of the $J/\psi$. The points represent the results obtained by HERA-B (black
    filled circles, with total and statistical uncertainties), at
    Fermilab~\cite{FNAL} and at the SPS~\cite{na50}. The data are fitted
    with linear functions.} 
  \label{fig:pt_vs_s}
  \end{center}
\end{figure}

\begin{figure}
  \begin{center}
  \includegraphics[width=0.4\textwidth]{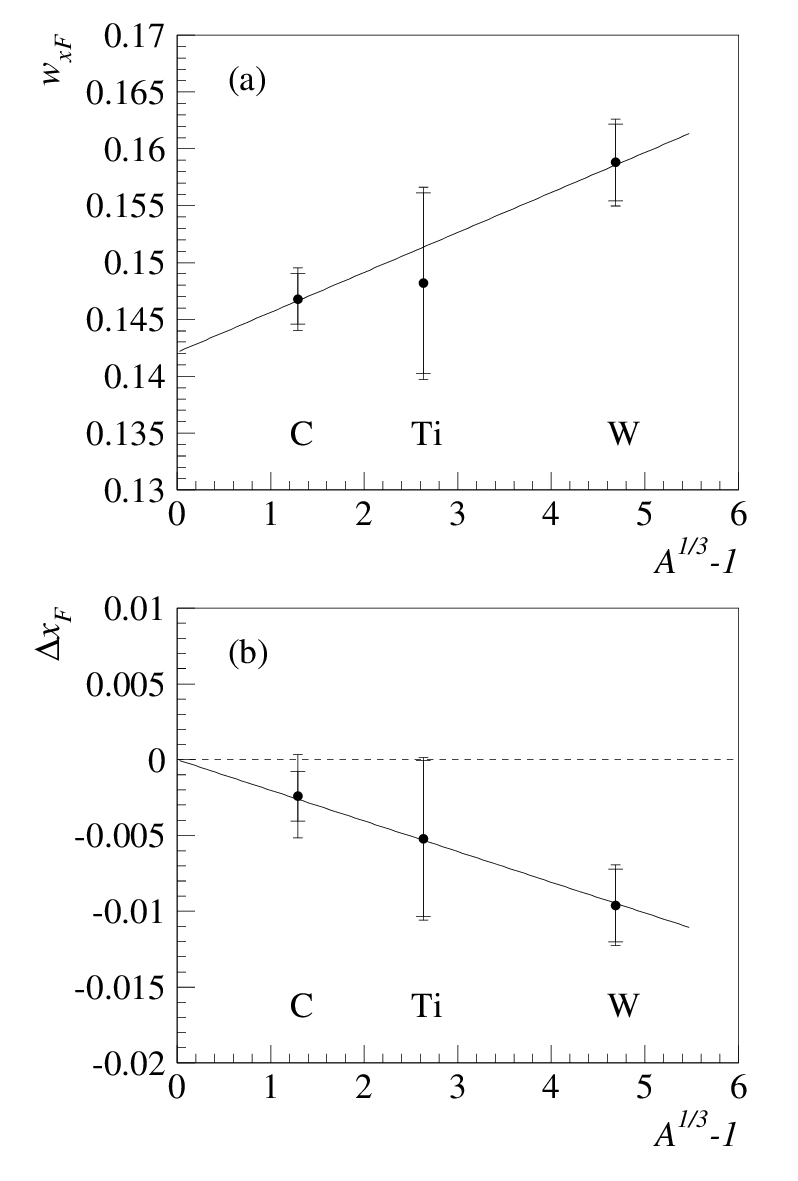}
  \caption{The width (a) and shift (b) of the $J/\psi$ $x_F$ distributions 
    as a function of $A^{1/3}-1$ (see Eqs.~\ref{eq:sigmaxF} and
    \ref{eq:deltaxF}). The double error bars represent
    total and statistical uncertainties. The data are fitted with linear
    functions.} 
  \label{fig:xf_vs_a13}
  \end{center}
\end{figure}

The dimuon and dielectron results (which, as discussed below, are found to be compatible) 
are averaged such that correlations in their systematic uncertainties are taken into 
account. Such correlations have been estimated by maintaining (when possible) a 
parallelism among the two channels when evaluating the effect of each single systematic 
test. The assumed polarization hypothesis is found to be the dominating source of uncertainty 
in the final results -- especially for the lower part of the $p_T$ distribution -- as well 
as the most important cause of correlation between the two analyses. Relative to this 
uncertainty, the signal selection cuts are in general responsible for negligible systematic 
variations, except for the most positive part of the $x_F$ spectrum, where acceptance 
corrections increase dramatically due to the low-angle detector acceptance cut-off near 
the beam.

The results are well represented -- and can therefore be described
-- by the following interpolating functions which are further motivated in the 
following section:
\begin{eqnarray}
  \frac{dN}{dp_{T}} & \ \propto \ & p_{T} \left(1 + \frac{1}{\beta - 2} \,
    \frac{p_{T}^{2}}{\langle p_{T}^{2} \rangle} \right) ^{-\beta},
  \label{eq:ptfit} \\
  \frac{dN}{dx_{F}} & \ \propto \ & \exp \left[ -\ln 2 \,
    \left|\frac{x_{F} - \Delta x_{F}}{w_{x_{F}}}\right|^{\gamma} \right].
  \label{eq:xffit}
\end{eqnarray}
The parameters $\langle p_T^2 \rangle$, $\beta$, $w_{x_F}$ (width at half maximum), 
$\Delta x_F$ (shift of the center of the distribution with respect to $x_F = 0$) 
and $\gamma$ are left free in the fit of the distributions. The resulting values 
are listed in Table~\ref{tb:distr_params} for each target material. The systematic
uncertainties of the parameters have been determined from the maximum variation 
(divided by $\sqrt{12}$) of the results obtained by re-fitting the distributions
after each of the stability tests described above. The normalized $\chi^{2}$ 
obtained from the fits of the combined dimuon-dielectron results are, respectively, 
$1.5$, $0.8$ and $2.1$ for the $p_T$ distributions of carbon, titanium and tungsten, 
and $1.1$, $1.5$ and $1.2$ for the three $x_F$ distributions (only the statistical 
uncertainties are taken into account).

Also included in Table~\ref{tb:distr_params} are the results obtained separately in 
the dimuon and dielectron channels with the respective systematic uncertainties. 
The good agreement between the results of the two analyses within the statistical 
uncertainties confirms that the channel-specific issues of particle identification 
and counting are not responsible for large systematic variations in the shapes 
of the distributions.

\subsection{Parameterization and interpretation of kinematic distributions} 
\label{sec:kindist}

Among the parameters adopted for the description of the data, the width of the 
$p_{T}$ distribution ($\langle p_{T}^{2} \rangle$), the position of the maximum 
of the $x_{F}$ distribution ($\Delta x_{F}$) and, possibly but less significantly, 
its width ($w_{x_{F}}$) show a trend with the mass number $A$.

It is well known~\cite{Cronin} that the average $p_{T}^{2}$ of particles produced 
in nuclear collisions increases with the mass of the target nucleus. This trend is
confirmed by the HERA-B data with high significance. The ``$p_T$-broadening'' effect 
is commonly explained as a consequence of multiple elastic scattering of
the incoming beam parton in the surrounding nucleus before the hard scattering process
takes place. The measured increase of $\langle p_{T}^{2} \rangle$ with $A$ is shown in 
Fig.~\ref{fig:pt2_vs_a13} together with the results of experiments at lower energies. 
The variable of the abscissa, $A^{1/3}-1$, is approximately proportional to the 
radius of the target nucleus i.e. to the average path length of the parton inside the 
nucleus with the shift of $-1$ such that the magnitude of the effect is 
measured with respect to $A=1$. All measurements are actually consistent with the 
parameterization
\begin{equation}
  \langle p_T^2 \rangle = \langle p_T^2 \rangle^{pp} + \rho \ (A^{1/3}-1), 
\label{eq:ptbroad}
\end{equation}
to which they are fitted in the plot. As summarized in
Fig.~\ref{fig:pt_vs_s}(a), the results for the average $p_{T}^{2}$
extrapolated to proton-nucleon interactions ($\langle p_{T}^{2}
\rangle^{pp}$) are compatible with a linear growth with the square of
the center-of-mass production energy $s$. On the other hand, $\rho$ is
approximately independent of center-of-mass energy as can be seen in
Fig.~\ref{fig:pt_vs_s}(b).

\begin{table}
    \begin{center}
    \begin{tabular}{l @{\hspace{5mm}}r@{.}l  @{\hspace{1mm}$\pm$\hspace{1mm}}r@{.}l @{\hspace{1mm}$\pm$\hspace{1mm}}r@{.}l    
}
      \multicolumn{7}{c}{$p_T$ distribution } \\ 
      \hline
      $\langle p_T^2 \rangle^{pp}$ ($\mathrm{GeV^2/c^2}$) &  2&030    & 0&014   & 0&014       \\
      $\rho   $  ($\mathrm{GeV^2/c^2}$)                   &  0&0852   & 0&0053  & 0&0043      \\
      $\beta^{pp}$                                        &  6&97     & 0&22    & 0&23        \\
      $\beta^{\prime}$                                    &  0&261    & 0&087   & 0&070       \\ 
      $\chi^2/NDoF$                                       &  \multicolumn{3}{c}{$106.5 / 71$}&\multicolumn{3}{c}{} \\
      \hline\hline 
      \multicolumn{7}{c}{$x_F$ distribution, $\tau$ free} \\ 
      \hline
      $\kappa$                                            &  -0&00211 & 0&00044 & 0&00022     \\
      $w_{x_F}^{pp}$                                      &   0&1435  & 0&0017  & 0&0046      \\
      $\tau$                                              &   0&00277 & 0&00066 & 0&00092     \\
      $\gamma^{pp}$                                       &   1&735   & 0&029   & 0&014       \\
      $\chi^2 / NDoF$                                     &  \multicolumn{3}{c}{$72 / 60$}&\multicolumn{3}{c}{}    \\
      \hline\hline 
      \multicolumn{7}{c}{$x_F$ distribution, $\tau = 0$ } \\  
      \hline
      $\kappa$                                            &  -0&00295 & 0&00035 & 0&00018     \\
      $w_{x_F}^{pp}$                                      &   0&1478  & 0&0014  & 0&0028      \\
      $\tau$                                              & \multicolumn{3}{c}{0 (fixed)}&\multicolumn{3}{c}{}     \\
      $\gamma^{pp}$                                       &   1&725   & 0&0028  & 0&0015      \\
      $\chi^2 / NDoF$                                     &  \multicolumn{3}{c}{$91.5 / 61$}&\multicolumn{3}{c}{}  \\\hline
    \end{tabular}

  \caption{Results of the global fits of $p_{T}$ and $x_{F}$
  distributions described in the text.  The first of the given
  uncertainty ranges is statistical and the second is systematic. }

  \label{tb:distr_params_global}
  \end{center}
\end{table}

Furthermore, HERA-B observes a difference in shape between the $x_{F}$ 
distributions of the $J/\psi$ for different target nuclei consisting of 
an increasing displacement of the center of the distribution towards negative 
values. As shown in Fig.~\ref{fig:xf_vs_a13}, $J/\psi$s are produced 
in tungsten with an $x_{F}$ distribution which has equal or slightly greater 
width with respect to those produced in carbon and tends to be asymmetrically 
centered at a lower value. This behavior is also supported by a fit of the E789 
gold data~\cite{FNAL} ($-0.035 < x_{F} <0.135$, $E_{b} = 800\,\mathrm{GeV}$) with 
Eq.~\ref{eq:xffit}. The fitted width of $0.11 \pm 0.01$ is lower than our value 
for tungsten suggesting not only that the maximum is shifted but that the shape 
becomes asymmetric. As a possible interpretation, the effect may be attributed 
to the energy loss undergone by the incident parton and/or the produced state in 
their path through the nucleus, causing a reduction of the average $x_F$ of the 
$J/\psi$ and a possible additional smearing of the momentum distribution. This 
hypothesis motivates the choice of representing the data also in this case as 
a function of $A^{1/3}-1$. The points in Fig.~\ref{fig:xf_vs_a13} (a) and
b) are fitted respectively with:
\begin{eqnarray}
  w_{x_F} &  =  &  w_{x_F}^{pp} + \tau \ (A^{1/3}-1), \label{eq:sigmaxF} \\
  \Delta x_F &  =  &  \kappa \ (A^{1/3}-1), \label{eq:deltaxF} 
\end{eqnarray}
where $\kappa$, $w_{x_F}^{pp}$ and $\tau$ are free parameters.

To obtain the best description of the dependence of the $p_{T}$ and $x_{F}$ 
spectra on the target nucleus, a simultaneous fit of the three (C, Ti and W) 
distributions according to the functions given in Eq.~\ref{eq:ptfit} and 
\ref{eq:xffit} has been done. According to the hypothesis that energy loss
is responsible for the observed nuclear dependence 
of the shape of the kinematic distributions, the fit has been constrained by 
imposing the relations from Eqs.~\ref{eq:ptbroad}, \ref{eq:sigmaxF} and 
\ref{eq:deltaxF}, and, moreover,
\begin{eqnarray}
  \beta & =  & \beta^{pp} + \beta^{\prime} (A^{1/3}-1), \label{eq:beta} \\
  \gamma & =  & \gamma^{pp} \textrm{ (independent of $A$)},
  \label{eq:gamma}
\end{eqnarray}
where $\beta^{pp}$, $\beta^{\prime}$ and $\gamma^{pp}$ are additional parameters of the 
fit. In Table~\ref{tb:distr_params_global} the results of this procedure are summarized.
The fit of the $x_{F}$ distributions has been performed in two variants, with the 
parameter $\tau$ left free or fixed to zero -- therefore assuming in
the latter case that $w_{x_F}$ is independent of $A$. The resulting best-fit curves 
(with $\tau$ left free) are the interpolating lines plotted in Figs.~\ref{fig:pt_ctiw} 
and \ref{fig:xf_ctiw}. The fit results indicate a significant nuclear dependence not only 
of the $p_{T}$ distribution (parameter $\rho$), but also of the $x_{F}$ distribution: 
there is a significance of $7 \sigma$ for $\kappa \neq 0$ when $\tau$ is fixed to zero, 
which changes to $4$ and $3 \sigma$, respectively, for $\kappa$ and $\tau$ 
(with a strong anti-correlation between the two) when both are left free.

\section{Nuclear dependence of $\boldmath{J/\psi}$ production} \label{sec:adep}
The Glauber Model~\cite{glauber_70} suggests that the dependence of
the $J/\psi$ production cross section on atomic mass number ($A$) can
be approximated by a power law:
\begin{equation}\label{eq:adep_pl}
  \sigma_{pA} = \sigma_{pN} \cdot A^{\alpha} \; ,
\end{equation}
where $\sigma_{pN}$ is the proton-nucleon cross section and $\alpha$,
the ``suppression'' parameter, characterizes the nuclear dependence.
Pure hard scattering in the absence of any nuclear effects would 
correspond to $\alpha$ equal to unity. A suppression of $J/\psi$
production would lead to $\alpha < 1$ while an enhancement
(anti-screening effect) would be signaled by $\alpha > 1$. Usually,
$\alpha$ is described and measured as a function of $x_F$ and $p_T$
(see for example~\cite{e866_adep,na50,herab_kstarphi}).

Eq.~\ref{eq:adep_pl} is generally used to describe data and
predictions independently of particular mechanisms of nuclear
modification. In general, however, $\alpha$ may depend on $A$ and thus
depend on the targets used to make the measurement.  For the
measurement presented here, $\alpha$ is evaluated by comparing the
$J/\psi$ yields from two different targets: carbon and
tungsten\footnote{
The titanium sample is not used for this analysis since it is too small 
to have a significant impact on the statistical precision of the result 
and would have required a significantly more complex analysis procedure.} 
as a function of $x_F$ and $p_T$.

\subsection{The $\alpha$ measurement}

Using Eq.~\ref{eq:adep_pl}, the nuclear suppression parameter $\alpha$
can be extracted from a measurement of the ratio of cross-sections for
carbon (C) and tungsten (W) targets:
\begin{equation}\label{eq:adep_pl2}
  \frac{\sigma_{pW}}{\sigma_{pC}} =
  \left(\frac{A_W}{A_C}\right)^\alpha. 
\end{equation}

The measurement of the cross section ratio requires a measurement of
the ratio of the integrated luminosities of the carbon and tungsten
target samples. For the HERA-B setup, this can be done using data
samples where two different targets are operated simultaneously
(double-target runs) since most of the systematic uncertainties cancel
and an absolute luminosity measurement can be avoided. On the other
hand, for studies of the dependence of $\alpha$ on the kinematic
variables, greater statistical precision can be obtained by also using
the single-target runs. The HERA-B measurement of $\alpha$ thus
consists of two sub-measurements: a measurement of the average value
of $\alpha$, $\langle \alpha \rangle$, over the full visible kinematic
range and a measurement of $\alpha - \langle \alpha \rangle$ as a
function of $x_F$ and $p_T$. The shape distributions
are then corrected using $\langle \alpha \rangle$ to produce an
absolute measurement of the distribution of $\alpha$ over the measured
range.

More specifically, $\langle \alpha \rangle$ is evaluated using double
target runs based on the formula:
\begin{equation} \label{eq:adep_alpha}
  \langle \alpha \rangle = \frac{\ln(\frac{\sigma_{\mathrm{W}}}{\sigma_{\mathrm{C}}})}
  {\ln(\frac{A_{\mathrm{W}}}{A_{\mathrm{C}}})} =
  \frac{1}{\ln(\frac{A_{\mathrm{W}}}{A_{\mathrm{C}}})} \cdot
  \ln\left(\frac{N_{\mathrm{W}}}{N_{\mathrm{C}}}
    \cdot \frac{\mathcal{L}_{\mathrm{C}}}{\mathcal{L}_{\mathrm{W}}}
    \cdot\frac{\epsilon_{\mathrm{C}}}{\epsilon_{\mathrm{W}}}\right)\, ,
\end{equation}
where $N_\mathrm{X}$ (X=C,W) denotes the total number of reconstructed
$J/\psi$ mesons originating from the corresponding target wire,
$\mathcal{L}_\mathrm{X}$ is the luminosity, $\epsilon_\mathrm{X}$
is the overall detection efficiency (see Sec.~\ref{sec:data}) and the
event yields are derived as discussed in Sec.~\ref{sec:analysis}. The
measurement of the luminosity ratios is described in
Sec.~\ref{sect:lumirat}.

The dependence of $\alpha$ on $x_{F}$ and $p_{T}$ is obtained from the
full carbon and tungsten target data samples (double- and single-target runs). 
The full carbon sample is roughly twice the size of the double-target 
subsample while the full tungsten sample is 10\% larger than the double-target 
subsample. The shape of the differential distributions are given by (here written
e.g. for $x_{F}$):
\begin{equation}\label{eq:adep_signorm}
  \frac{1}{\sigma_{J/\psi}} \cdot \frac{d \sigma_{J/\psi}}{d \, x_{F}}\, ,
\end{equation}
where $\sigma_{J/\psi} =\sigma(pA \rightarrow J/\psi + X)$ is the total visible
$J/\psi$ cross section.

The measurement of nuclear effects can then be derived from the distributions 
of Eq.~\ref{eq:adep_signorm} 
using (here written e. g. for $x_{F}$):
\begin{equation}\label{eq:adep_alphaminus}
  \alpha(x_{F}) = 
  \frac{1}{\ln\Big(\frac{A_{\mathrm{W}}} {A_{\mathrm{C}}}\Big)}
  \cdot \ln\left(\frac
    {\frac{1}{\sigma_{W}} \cdot \frac{d \sigma_{W}}{d \, x_{F}}} 
    {\frac{1}{\sigma_{C}} \cdot \frac{d \sigma_{C}}{d \, x_{F}}} \right)
  + \langle \alpha \rangle \; .
\end{equation}

\subsection{Luminosity ratios} \label{sect:lumirat}

The luminosity acquired on target $\mathrm{X}$, where X is either C
(carbon) or W (tungsten), can be expressed as:
\begin{equation}
  \mathcal{L}_\mathrm{X} = \frac{N_{\mathrm{X}}}{\sigma^{\mathrm{inel}}_\mathrm{X}} =
  \frac{\lambda_\mathrm{X} \cdot N^{BX}}{\sigma^{\mathrm{inel}}_\mathrm{X}}\,,
\end{equation}
where $N_{\mathrm{X}}$ is the total number of inelastic interactions occurring
on target $\mathrm{X}$ during the measurement,
$\sigma^{\mathrm{inel}}_\mathrm{X}$ is the total inelastic cross section,
$N^{BX}$ is the corresponding total number of filled bunch crossings
(BX) and $\lambda_\mathrm{X}$ is the average number of interactions
per filled BX. The total cross sections $\sigma^{\mathrm{inel}}_\mathrm{X}$ 
for each target material together with some details on this topic can be 
found in~\cite{lumi_HB}. The luminosity ratio $R_{\mathcal{L}}$ needed
in Eq.~\ref{eq:adep_alpha} is then given by:
\begin{equation} \label{eq:lr_def}
  R_{\mathcal{L}} = \frac{\mathcal{L}_\mathrm{C}}{\mathcal{L}_\mathrm{W}} =
  \frac{\sigma^\mathrm{inel}_\mathrm{W}}{\sigma^\mathrm{inel}_\mathrm{C}} 
  \cdot 
  \frac{\lambda_\mathrm{C}}{\lambda_\mathrm{W}} \; .
\end{equation}

Assuming the interaction probability on target X follows a Poisson
distribution, $\lambda_X$ can be calculated from the observed number
of events with at least one interaction ($N(\geq 1)^{\mathrm{obs}}$)
using:
\begin{equation}  \label{eq:lambda}
  \lambda_X =  \frac{1}{\epsilon_X^{inel}}
  \ln\Bigg(1 - \frac{N(\geq 1)^{\mathrm{obs}}}{N^{\mathrm{BX}}}\Bigg)\; ,
\end{equation}
where $\epsilon_\mathrm{X}^{inel}$ is the probability to observe a single interaction.

The determination of $\lambda_{\mathrm{X}}$ relies on random-trigger
events which were accumulated together with the dilepton-trigger
events used for $J/\psi$ counting.  Five methods which
differ by the event characteristics used to define the presence of an
interaction are used to count events.  All methods rely on tracks
found in the vertex detector.  To maintain high efficiency, the
requirements imposed are minimal but sufficient to also keep the
probability of incorrect target wire assignment at a low level.  The methods
are based on the following five criteria:
\begin{itemize}
  \item[] For all events,
    \begin{itemize}
    \item[1.] $\geq 1$ primary vertex on wire X where the primary
      vertex is formed from tracks measured both in the VDS
      and OTR (``long tracks''),
    \item[2.] $\geq 2$ tracks (including long tracks and tracks seen
      only in the VDS) with impact parameter $\leq 3 \sigma$ of wire
      X and $\geq 5 \sigma$ from the other wire, where $\sigma$ is the
      impact parameter measurement uncertainty,
    \end{itemize}
  \item[] and, using the subset of events with no vertex found on the
    other wire,
    \begin{itemize}
      \item[3.] $\geq 1$ primary vertex on wire X using all tracks,
      \item[4.] $\geq 1$ primary vertex on wire X using long tracks only,
      \item[5.] $\geq 2$ long tracks within $\leq 3 \sigma$ from wire X.
    \end{itemize}
\end{itemize}

Both as a cross check and as an estimate of the systematic uncertainty on
the efficiency, all counting methods are checked in parallel. For the
final luminosity ratio determination, the average of the five
determinations is used and the rms spread of the five is factored into
the systematic uncertainty.  Furthermore, Eq.~\ref{eq:lambda} assumes that 
the interaction probabilities for each wire follow a Poisson distribution.  
However, the individual bunch fillings are often uneven and the 
interaction rate varies in time by typically 20\%. To quantify this 
influence, an alternative luminosity calculation is performed in which the 
detailed bunch filling structure and the interaction rate distribution 
on each wire are taken into account. The differences between the
resulting ratios and those computed directly from Eq.~\ref{eq:lr_def}
are negligible compared to other systematic uncertainties.

To minimize the dependence of the efficiency estimate on MC, the
efficiency of each of the above methods is calibrated by comparing the
luminosity estimate found using it with that found by the methods
described in~\cite{lumi_HB}.  These latter methods rely on very
simple criteria to identify events with interactions, such as a
minimal number of hits in the RICH detector (typically twenty,
compared to thirty hits expected for a fully accepted fast charged
particle) or a small (1\,GeV) energy deposit in the ECAL and are estimated
to be sensitive to roughly 95\% of the total non-diffractive cross
section.

Each target of a two-wire configuration is calibrated separately using
single-wire data runs taken nearby in time to the run being
calibrated. A ``ghost'' wire is introduced at the location of the
other wire of the configuration.  Thus the MC is not relied on to
model tracking in the VDS or vertex finding, but only to estimate the
efficiency of simpler and more robust event counting techniques
described in~\cite{lumi_HB}.  Overall efficiencies in the range of 60
-- 80 \% are found, depending on method and wire. The
efficiency-calibration method based on real data is also used to
evaluate the probability that interactions are assigned to the wrong
wire. This probability is method and configuration dependent and is
never more than 0.4\%.

The average relative systematic uncertainties on the luminosity ratios due to 
interaction counting and MC calibration~\cite{lumi_HB} are 1.3\% and 3.2\%,
respectively, giving an overall scale uncertainty of 3.4\% on $R_{\mathcal{L}}$. 
Depending on wire configuration between 0.6 and 1.2 million events were used 
for the determination of $\lambda_\mathrm{C}$ and $\lambda_\mathrm{W}$, thus
the statistical uncertainty on the luminosity ratio is negligible.

\subsection{Results}

Based on Eq.~\ref{eq:adep_alpha}, an average suppression value of
\begin{equation} \label{eq:alpha_mean}
  \langle \alpha \rangle = 0.981 \pm 0.004_{\mathrm{stat.}} \pm 0.016_{\mathrm{sys.}}
\end{equation}
in the visible range of $x_{F}$ is obtained. As explained in
Sect.~\ref{sec:data}, the target system of HERA-B consisted of eight
different wires grouped in two stations. The data were taken with four
different two-wire configurations which were analyzed separately and
averaged to obtain the final value. Also, the electron and the muon
decay channels represent two statistically independent measurements. 
The average value is calculated as a weighted mean with weights being 
the squared quadratic sum of statistical and luminosity ratio 
uncertainties. The luminosity ratios are determined for each wire 
configuration separately as discussed in Sec.~\ref{sect:lumirat}
with a contribution to the systematic uncertainty on $\langle \alpha
\rangle$ of $3.4~\% / \ln(A_{W}/A_{C}) = 1.24~\%$.  A systematic effect 
of 1.1\% due to time variatons of detector performance and imprecisions 
in detector or trigger simulations was estimated from the variations of 
$\langle \alpha \rangle$ among the four samples and two decay channels.  
The total systematic uncertainty on $\langle \alpha \rangle$ is thus 
$1.66~\%$. The statistical precision of the $\langle \alpha \rangle$ 
measurement contributes an uncorrelated uncertainty $0.4~\%$.  

The values of $\alpha$ for individual $x_{F}$ and $p_{T}$ bins are 
given in Table~\ref{tb:adep} and shown in Fig.~\ref{fig:adep_vs_pt}
for $p_{T}$ and in Fig.~\ref{fig:adep_vs_xf} for $x_{F}$. They 
will be further discussed in Sec.~\ref{sec:interp}. The values
and systematic uncertainty estimates were derived according to the
procedure described in Sec.~\ref{sec:kine_results}. 
The error bars on the figures show 
both statistical and total contributions. The systematic uncertainties
in the estimate of $\langle \alpha \rangle$ are largely uncorrelated
with those from the $\alpha - \langle \alpha \rangle$ measurement.
The final systematic uncertainty estimate is the quadratic sum of the
two.  The systematic uncertainty is substantially correlated from bin to
bin.

\subsection{Discussion} \label{sec:interp}

\begin{figure}
  \begin{center}
  \includegraphics[width=0.47\textwidth]{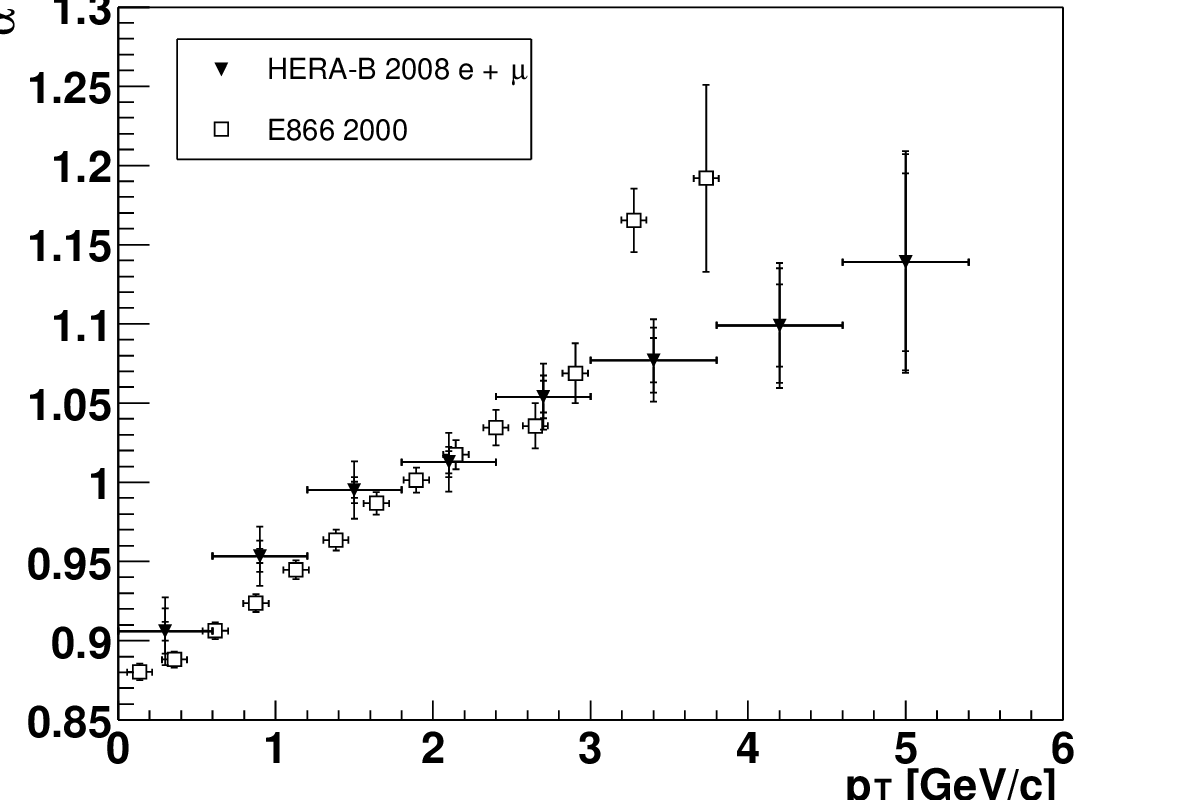}
  \caption{The nuclear suppression parameter $\alpha$ as a function of $p_{T}$
    measured by HERA-B (filled triangles, plotted with total and statistical uncertainties)
    and by E866~\cite{e866_adep} (empty squares).} 
  \label{fig:adep_vs_pt}
  \end{center}
\end{figure}

\begin{figure*}
  \begin{center}
  \includegraphics[width=0.95\textwidth]{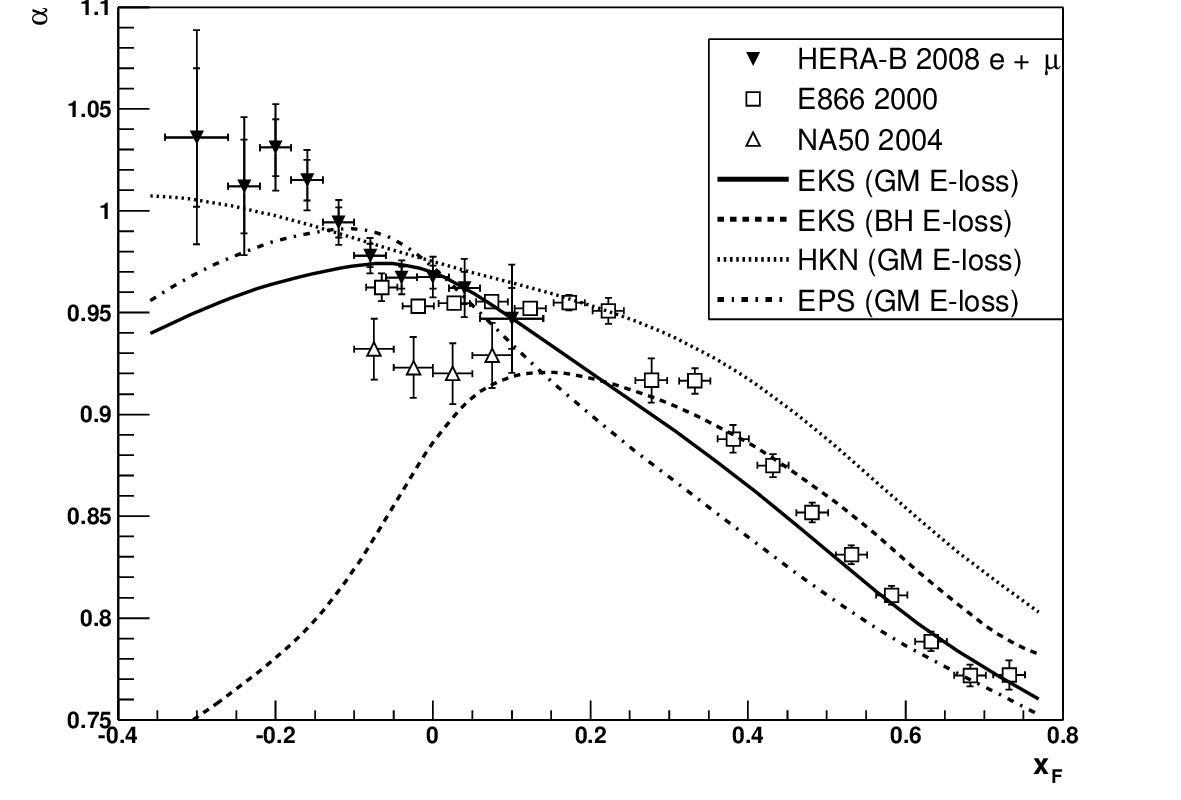}
  \caption{Measurements of $\alpha$ as a function of $x_F$ by HERA-B
  (filled triangles, plotted with total and statistical
  uncertainties), E866 ($\sqrt{s} =
  38.8\,\mathrm{GeV}$)~\cite{e866_adep} (empty squares) and NA50
  ($\sqrt{s} = 29.0\,\mathrm{GeV}$)~\cite{na50} (empty triangles).
  The curves were calculated by Vogt~\cite{vog_00,vog_05} based on
  three different nuclear parton distribution functions:
  (EPS~\cite{EPS_08}, EKS~\cite{EKS_98,EKR_98} and HKN~\cite{HKN_04})
  and two models of initial state energy-loss: GM~\cite{gavin_mi} and
  BH~\cite{brodsky_h}. For all approaches, energy loss, intrinsic charm
  and shadowing are taken into account.}  \label{fig:adep_vs_xf}
  \end{center}
\end{figure*}

The results of the $\alpha$ measurement as functions of $p_{T}$ and $x_{F}$ 
are given in Table~\ref{tb:adep}. The distributions of the $\alpha(p_{T})$ 
and $\alpha(x_{F})$ are presented in Figs.~\ref{fig:adep_vs_pt} and 
\ref{fig:adep_vs_xf} where they are compared with measurements performed 
by other fixed target experiments: E866~\cite{e866_adep} ($E_{p} =
800\,\mathrm{GeV}$) and NA50~\cite{na50} ($E_{p} =
450\,\mathrm{GeV}$). As already seen (e.g. in
Fig.~\ref{fig:pt2_vs_a13}), the measured $p_{T}$ dependence of the
nuclear modification effects is very similar for HERA-B and E866. In
Fig.~\ref{fig:adep_vs_xf} the E866 and HERA-B measurements are seen to
be compatible within statistical and systematic uncertainties in the
overlap region. The NA50 results are based on lower energy collisions and
are systematically below both HERA-B and E866.  At lower values of
$x_F$, the HERA-B $\alpha(x_{F})$ measurement indicates a reversal of
the suppression trend seen at high $x_F$: the strong suppression
established by previous measurements at high $x_{F}$ turns into a
slight tendency towards enhancement in the negative $x_{F}$ region.

\begin{table}
    \begin{center}
    \begin{tabular}{
                    r@{.}l 
                    r@{.}l|
                    @{\hspace{7mm}}r@{.}l  @{\hspace{1mm}$\pm$\hspace{1mm}}r@{.}l @{\hspace{1mm}$\pm$\hspace{1mm}}r@{.}l    }
       \multicolumn{4}{c|@{\hspace{7mm}}}{$p_T$ ($\mathrm{GeV/c}$)}
      &\multicolumn{6}{c}{ $\alpha$}                    \\
       \multicolumn{2}{c}{min }
     & \multicolumn{2}{c|@{\hspace{7mm}}}{max } 
     & \multicolumn{6}{c}{}                             \\
      \hline
      0&0 & 0&6     & 0&906 & 0&006 & 0&021 \\
      0&6 & 1&2     & 0&953 & 0&005 & 0&019 \\
      1&2 & 1&8     & 0&995 & 0&005 & 0&017 \\
      1&8 & 2&4     & 1&013 & 0&007 & 0&018 \\
      2&4 & 3&0     & 1&054 & 0&010 & 0&019 \\
      3&0 & 3&8     & 1&077 & 0&014 & 0&021 \\
      3&8 & 4&6     & 1&099 & 0&026 & 0&028 \\
      4&6 & 5&4     & 1&139 & 0&056 & 0&038 \\
      \hline\hline
       \multicolumn{4}{c|@{\hspace{7mm}}}{$x_F$}
      &\multicolumn{6}{c}{ $\alpha$}                    \\
       \multicolumn{2}{c}{min }
     & \multicolumn{2}{c|@{\hspace{7mm}}}{max } 
     & \multicolumn{6}{c}{}                             \\
      \hline
      -0&34 & -0&26 & 1&036 & 0&034 & 0&042 \\
      -0&26 & -0&22 & 1&012 & 0&023 & 0&030 \\
      -0&22 & -0&18 & 1&031 & 0&014 & 0&023 \\
      -0&18 & -0&14 & 1&015 & 0&010 & 0&020 \\
      -0&14 & -0&10 & 0&994 & 0&008 & 0&018 \\
      -0&10 & -0&06 & 0&978 & 0&006 & 0&018 \\
      -0&06 & -0&02 & 0&967 & 0&005 & 0&017 \\
      -0&02 &  0&02 & 0&967 & 0&006 & 0&018 \\
       0&02 &  0&06 & 0&962 & 0&008 & 0&021 \\
       0&06 &  0&14 & 0&947 & 0&015 & 0&028 \\ \hline
    \end{tabular}
  \caption{The parameter $\alpha$ as a function of $p_{T}$ and $x_{F}$.
    Statistical and systematic uncertainties are indicated separately. They
    were extracted from the measurement using Eq.s~\ref{eq:adep_alphaminus}
    and \ref{eq:alpha_mean}.}
  \label{tb:adep}
  \end{center}
\end{table}

The dependence of $J/\psi$ production in hadron-nucleus interactions
on $x_{F}$ has been modeled by Vogt~\cite{vog_00,vog_05}. Nuclear
effects caused by final-state absorption, interactions with co-movers,
shadowing of parton distributions, energy loss and intrinsic charm
quark components are described separately and integrated into the
model. It is further assumed that the $c\bar{c}$ pair is subject to
more severe energy losses if produced in a color octet state.  Four
curves from this model which differ in their descriptions of nuclear
Parton Density Functions (nPDF) and energy loss are shown in
Fig.~\ref{fig:adep_vs_xf}. All calculations shown here were done for
the center-of-mass energy of HERA-B at $\sqrt{s}
=41.6\,\mathrm{GeV}$. The nPDF distributions of Eskola, Kolhinen and
Salgado (EKS)~\cite{EKS_98} describe the scale dependence of the
ratios of nPDFs of a proton inside a nucleus to those of a free proton
within the framework of lowest order leading-twist DGLAP
evolution~\cite{DGLAP} by evolving the initial PDF from the
CTEQ4L~\cite{CTEQ4L} and leading order GRV~\cite{GRV_LO}
parameterizations. An improved leading-order DGLAP analysis of nPDFs
including next to leading order calculations has been published
recently by Eskola, Paukkunen and Salgado (EPS)~\cite{EPS_08}. In
another approach by Hirai, Kumano and Nagai (HKN)~\cite{HKN_04},
nuclear structure function ratios
$F_{2}^{\mathrm{A}}/F_{2}^{\mathrm{A}^{\prime}}$ and Drell-Yan cross
section ratios are analyzed to obtain nPDFs. The HKN analysis shows
weak anti-shadowing at negative $x_{F}$.

Initial state energy loss as described by Gavin and Milana (GM)~\cite{gavin_mi} 
and modified by Brodsky and Hoyer (BH)~\cite{brodsky_h} 
is based on a multiple scattering approach that essentially depletes the 
projectile parton momentum fraction as the parton moves through the nucleus. 
Both quarks and gluons can scatter elastically and therefore lose energy prior to 
the hard process resulting in an effective reduction of $J/\psi$ production 
for $x_{F} > 0$.

The measurement of HERA-B shows that $\alpha$ increases with decreasing $x_F$ and 
suggests enhanced $J/\psi$ production for $x_F < -0.1$.  The HERA-B data favors 
the nPDFs of EPS and HKN over EKS. The BH description of energy loss is clearly 
ruled out.  None of the variants of the Vogt model give a satisfactory description 
of both HERA-B and E866 data. For example, while the HKN curve is compatible
with most of the HERA-B data points at negative $x_{F}$, it lies significantly above 
the E866 points and furthermore fails to adequately describe RHIC data~\cite{EPS_08}.

Another theoretical model by Boreskov and Kaidalov~\cite{borkaid},
formulated in the framework of reggeon phenomenology, predicts an
anti-screening effect in the region of negative $x_{F}$. An important
ingredient of their model is the assumption that a colorless state
containing $c$ and $\bar{c}$ quarks which has some probability of
projecting into a charmonium state is produced and propagates through
the nucleus. The colorless state is of large size, possibly consisting
of $D\bar{D}$ or $D^{\ast}\bar{D}^{\ast}$ mesons, and therefore has a
large interaction cross section. As it propagates through the nucleus
it interacts and loses energy.  Ultimately, the observed $J/\psi$
mesons are projected out of the energy-depleted colorless state.  The
measurements of HERA-B are qualitatively compatible with the
calculations described in BK~\cite{borkaid}.

\section{Conclusions} \label{sec:conclusions}
HERA-B has performed the first determination of the nuclear dependence
of $J/\psi$ production kinematics at negative $x_F$ in
proton-nucleus collisions. The analyzed data samples were obtained in
collisions of protons from the 920~\,GeV HERA-proton beam with carbon,
tungsten and titanium targets.  The $J/\psi$ mesons are observed in
both dimuon and dielectron decay channels. The comparison of results
from the two channels affords some additional control over systematic
uncertainties arising from triggering and selection procedures.

The measurement covers the kinematic range $-0.34 < x_F < 0.14$ and
$p_{T} < 5.4\,\mathrm{GeV/c}$. The measured $dN/dp_T$ distribution is seen to
become broader with increasing atomic mass number as has already been
observed by experiments at lower center-of-mass energies~\cite{FNAL,na50}. 
The data indicates that the $dN/dx_F$ distribution also tends to become 
broader and that its center moves towards negative $x_F$ values with 
increasing A.

The dependences of the nuclear suppression parameter, $\alpha$, on
$p_T$ and $x_F$ are also presented. The $\alpha$ parameter is seen to
increase with increasing $p_T$ in agreement with data from
E866~\cite{e866_adep}. In the $x_F$ region of overlap of the two
experiments, the two $\alpha$ measurements are mutually consistent. As
$x_F$ decreases, $\alpha$ increases and becomes greater than 1 below
$x_F \approx -0.15$, although the data remains compatible with a value
of 1 to within $2\,\sigma$.  Thus instead of the strong suppression
observed at high positive $x_F$, HERA-B measured no suppression or a
possible enhancement of $J/\psi$ production at negative
$x_F$. Hard-scattering based models~\cite{vog_00,vog_05} have
difficulty simultaneously accommodating the HERA-B, E866 and RHIC
measurements, while the reggeon-inspired model of Boreskov and
Kaidalov~\cite{borkaid} is in qualitative agreement with the data.

\section*{Acknowledgments}
We express our gratitude to the DESY laboratory for the strong support
in setting up and running the HERA-B experiment. We are also in debt
to the DESY accelerator group for their continuous efforts to provide
good and stable beam conditions. The HERA-B experiment would not have
been possible without the enormous effort and commitment of our
technical and administrative staff. It is a pleasure to thank all of
them. We thank R. Vogt for the predictions shown in
Fig.~\ref{fig:adep_vs_xf} and for useful discussions and guidance.

\end{document}